\documentclass[10pt, conference, letterpaper]{IEEEtran}

\ifCLASSINFOpdf
  \usepackage[pdftex]{graphicx}
  \DeclareGraphicsExtensions{.pdf,.jpeg,.png}
\else
\fi

\makeatletter
\newcommand\footnoteref[1]{\protected@xdef\@thefnmark{\ref{#1}}\@footnotemark}
\makeatother

\usepackage{amsmath}
\usepackage{array}
\usepackage{amssymb}
\usepackage{enumitem}
\usepackage{url}
\usepackage{textcomp}
\usepackage{gensymb}
\usepackage{siunitx}
\usepackage{rotating}
\usepackage{multirow}
\usepackage{xspace}
\usepackage{xcolor,color}
\usepackage{blkarray}
\usepackage{multirow}
\usepackage{graphicx}
\usepackage{epstopdf}
\usepackage{wrapfig}
\usepackage{xspace}
\usepackage[utf8]{inputenc}
\usepackage[font=small,skip=5pt]{caption}
\usepackage{subcaption}
\usepackage{mathtools}
\usepackage{mathptmx}
\usepackage[T1]{fontenc}
\usepackage{lmodern}
\usepackage{supertabular}
\usepackage{adjustbox}
\usepackage{times}
\usepackage{etoolbox}

\newcommand{\red}[1]{\textcolor{red}{#1}}

\begin{document}
	
\title{Analyzing Mobility-Traffic Correlations in Large WLAN Traces: \textit{Flutes} vs. \textit{Cellos}}

\makeatletter
\patchcmd{\@maketitle}
  {\addvspace{0.5\baselineskip}\egroup}
  {\addvspace{-1.5\baselineskip}\egroup}
  {}
  {}
\makeatother

\author{
\IEEEauthorblockN{~Babak Alipour\IEEEauthorrefmark{1}\\ \tt\footnotesize babak.ap@ufl.edu} \and 
\IEEEauthorblockN{Leonardo Tonetto\IEEEauthorrefmark{2}\\ \tt\footnotesize tonetto@in.tum.de} \and
\IEEEauthorblockN{Aaron Yi Ding\IEEEauthorrefmark{2}\\ \tt\footnotesize ding@in.tum.de}
\and 
\IEEEauthorblockN{Roozbeh Ketabi\IEEEauthorrefmark{1}\\ \tt\footnotesize roozbeh@ufl.edu} \and
\IEEEauthorblockN{Jörg Ott\IEEEauthorrefmark{2}\\ \tt\footnotesize ott@in.tum.de} \and
\IEEEauthorblockN{Ahmed Helmy\IEEEauthorrefmark{1}\\ \tt\footnotesize helmy@ufl.edu}
\and
\IEEEauthorblockA{
\IEEEauthorrefmark{1}Computer and Information Science and Engineering\\
University of Florida, Gainesville, USA\\
} \and
\IEEEauthorblockA{
\IEEEauthorrefmark{2}Department of Informatics\\
Technical University of Munich, Munich, Germany\\
}
}

\maketitle
\thispagestyle{plain}
\pagestyle{plain}

\begin{abstract}

Two major factors affecting mobile network performance are \textit{mobility}
and \textit{traffic} patterns. Simulations and analytical-based performance evaluations
rely on models to approximate factors affecting the network. Hence, the understanding
of mobility and traffic is imperative to the effective evaluation
and efficient design of future mobile networks.
Current models target either mobility or traffic, but do not capture their interplay. Many trace-based mobility models have largely used pre-smartphone datasets (e.g., AP-logs), or much coarser granularity (e.g., cell-towers) traces. This raises questions regarding the relevance of existing models, and motivates our study to revisit this area.
In this study, we conduct a multi-dimensional analysis, to \textit{quantitatively} characterize
mobility and traffic spatio-temporal patterns, for laptops and smartphones, leading
to a detailed integrated mobility-traffic analysis. Our study is \textit{data-driven},
as we collect and mine capacious datasets (with \textbf{30TB}, 300k devices)
that capture all of these dimensions. The investigation is performed using our systematic
(\textit{FLAMeS}) framework. Overall, dozens of mobility and traffic features
have been analyzed. The insights and lessons learnt serve as guidelines and
a first step towards future \textit{integrated mobility-traffic models}. In addition,
our work acts as a stepping-stone towards a richer, more-realistic
suite of \textit{mobile test scenarios} and \textit{benchmarks}.

\end{abstract}

\IEEEpeerreviewmaketitle

\section{Introduction}

Human mobility has been studied extensively and many models have been
derived.  The spectrum ranges from simple synthetic mobility models to
complex trace-based models, capturing different properties with
varying degrees of accuracy \cite{treuniet:csur14, hess:csur16}.
Similarly, network traffic has been studied 
increasingly for wireless networks: for rather stationary users (as in
WLANs) (e.g., \cite{kotz:campus-wifi05, henderson:wlan-usage08}) and
potentially mobile users as for cellular networks (e.g.,
\cite{maier:pam10,zhang:cellnet12}).  Such analyses range from
metrics such as flow count, sizes, and traffic volume to service usage
(e.g., visited web sites, backend services).

Both mobility and network usage, characterize different
aspects of human behavior. In this sense, we have a \textit{mobility
plane} and a \textit{(network) traffic plane}. In reality,
these two planes are likely interdependent.
Human mobility may be
influenced by network activity; for example, a person slowing down to
read incoming messages.  Also, network activity may be influenced by
mobility and location; stationary users may produce/consume more data
than those walking, and people may use
different services in different places \cite{Moghaddam2011}.

In earlier studies, this interdependence has not been widely
considered, and models for both mobility and network traffic planes
have been developed and evaluated largely in isolation.  For example,
when evaluating mobile systems’ performance, traffic generation
generally follows regular patterns, drawn from common simple
distributions (e.g., exponential or uniform), while assuming neither transmission
nor reception of data impacts mobility.  Simply observing people
walking while staring at (or reacting to) their smartphones suggests,
however, that such interdependencies need to be captured
properly. Understanding the mobility-traffic interplay is imperative
to the effective evaluation and efficient design of future mobile
algorithms ranging from user behavior prediction
and caching,
to network load estimation and resource allocation.

In this paper, we take a stab at understanding the interconnection of
the mobility and traffic planes.  To do this properly, we need to
consider the nature of mobile devices people use: one class of devices
is merely intended for stationary use, typically while the user is
seated---this primarily holds for laptop computers, dubbed
\textit{\textbf{cellos}}.  In contrast, another class---'on-the-go' smartphones, which we
refer to as \textit{\textbf{flutes}}---lend themselves to truly mobile
use\footnote{Throughout, we use \textit{flutes} for \textit{smartphones}, and \textit{cellos} for \textit{laptops}.}.  
We focus our analysis on these two classes
because they have been around long enough to have extensive datasets
to build upon.
We stipulate that the interconnection of the mobility and traffic is
modulated by the device(s) a mobile user is carrying. Therefore, we
follow two main lines of investigation: we develop a framework to
differentiate between cellos and flutes, and study both the mobility
and traffic patterns for each of those types.

Specifically, the main goal of this paper is to quantitatively investigate
the following questions in-depth: \textit{
(I) How different are mobility and traffic characteristics across device types, time and space?
(II) What are the relationships between these characteristics?
(III) Should new models be devised to capture these differences? And, if so, how?}

To answer these questions, a multi-dimensional (comparative) analysis
approach is adopted to investigate mobility and traffic
spatio-temporal patterns for flutes and cellos.  We drive our
study with capacious datasets (30TB+) that capture all the above
dimensions in a campus society, including over 300k devices (Sec.
\ref{sec:experiments}).  A systematic \textit{F}ramework for
\textit{L}arge-scale \textit{A}nalysis of \textit{M}obil\textit{e}
\textit{S}ocieties (\textit{FLAMeS}) is devised for this study, that
can also be used to analyze other multi-sourced data in future
studies.
%
%
%
Our main contributions include:

\begin{enumerate}

  \item \textit{Integrated mobility-traffic analyses} (Sec.
    \ref{sec:multi_dim}): This study is the first
    to quantify the correlations of numerous features of mobility and
    traffic simultaneously. This can identify gaps in existing mobile
    networking models, and reopen the door for future impactful work
    in this area.

  \item \textit{Flutes vs. Cellos analysis} (Sec.
    \ref{sec:mobility}--\ref{sec:traffic_analysis}): The device type
    classification presented here, is an important
    dimension to understand. This is particularly relevant as new
    generations of portable devices are introduced, that are different
    than laptops, traditionally considered in earlier studies.

  \item \textit{Systematic multi-dimensional investigation framework} (Sec.
    \ref{sec:flames}): \textit{FLAMeS} provides the scaffolding needed
    to process, in multiple dimensions, many features of large sets of
    measurements from wireless networks, including AP-logs and NetFlow
    traces. This systematic method can apply to other datasets in
    future studies.
  
\end{enumerate}


\section{Related work}
\label{sec:related}

To characterize mobility and network usage, existing studies have
covered various aspects, including human mobility, device variation,
and dataset analysis.
 
\textbf{Human Mobility}: 
Given its importance in various research areas, human
mobility has received significant attention. We refer 
the reader to \cite{treuniet:csur14,hess:csur16} for surveys of mobility
modeling and analysis.
For spatial-temporal patterns,
\cite{gonzalez2008understanding} and \cite{song2010modelling} reveal the 
regularity and bounds for predicting human mobility using cellular logs.
A recent study  highlighted the importance of
combining different datasets to study various features simultaneously \cite{Zhang2014}.
Our observations are similar to \cite{gonzalez2008understanding, song2010modelling},
which reaffirms the intrinsic properties of human mobility, despite differences in 
granularity and population across datasets.
To advance the understanding of human mobility, we integrated different datasets
to correlate mobility and network traffic.

\textbf{Device Variation}: 
Usage and traffic patterns of different device types have been studied from various
perspectives (\cite{maier2010first,Kumar2013,chen2012network,gember2011comparative,afanasyev2008analysis,papapanagiotou2012smartphones}).
However, those findings are based on classifications that
rely on either MAC addresses or HTTP headers solely.
The former is rather limited and the latter may have serious privacy implications
and are often unavailable.
In \cite{Falaki2010}, authors use packet-level traces from 10 phones and application-level monitoring from 33 Android devices to analyze smartphone traffic. Although this allowed fine-grained measurements, the approach is invasive and
limited in scalability, leading to small sample sizes and restricted conclusions.
They also do not compare the traffic of smartphones with that of ``stop-to-use''
wireless devices (i.e. cellos) nor do they measure spatial metrics.
The study in \cite{Das:TOIT2016} analyzes 32k users on campus, and focuses on multi-device usage. It notes differences between laptops and smartphones in packets, content, and time of usage. That work targets device usage patterns and security, while we study mobility and wireless traffic correlations.
In our method, the combination of MAC and NetFlow
allowed us to classify majority of observed devices while preserving users' privacy.

\textbf{Dataset Analysis}:
A recent work on WLAN traces \cite{cao:infocom2017} revealed
surprising patterns on increases of long-term mobility entropy 
by age, and the impact of academic majors on students' long-term mobility entropy.
The authors of \cite{Moghaddam2011} investigated correlations and characteristics
of web domains accessed by users and their locations, based on NetFlow
and DHCP
logs from a university campus in 2004. They propose a simulation paradigm with 
data-driven parameters, producing realistic scenarios for simulations. However,
that study uses data from pre-smartphone era and does not distinguish between
device types. It also does not analyze the relation between mobility and
traffic. 
On both WiFi and cellular networks, the authors of \cite{Nikravesh2016} performed an in-depth study on smartphone traffic, highlighting the benefits and limitations of using MPTCP. 
Distributions of flow  inter-arrival time (IAT) and arrival rate at APs of  ``static''
flows were analyzed (e.g., Exp, Weibull,
Pareto, Lognormal) in \cite{Meng2004}. Lognormal was found to best fit
the flow sizes, while at small time scales (i.e. hourly), IAT was best described
by Weibull but parameters vary from hour to hour.
We analyze flows on a much larger scale, newer dataset including smartphones,
and identify Lognormal distribution as the best fit for flow
sizes, and beta as best for IAT, regardless of device type.
The study in \cite{Xu:ToN2017} analyzed ISP traces with 9600 cellular towers and 150K users in Shanghai, and mapped timed traffic patterns to urban regions. It provided insight into mobile traffic patterns across time, location and frequency. This work is complementary to ours, as we provide a much finer scope analyzing campus WLAN traces.

\section{Systematic multi-dimensional analysis} 
\label{sec:flames}

\begin{figure}[ht]
    \centering
    \setlength{\belowcaptionskip}{-5pt}
    \includegraphics[width=0.98\linewidth]{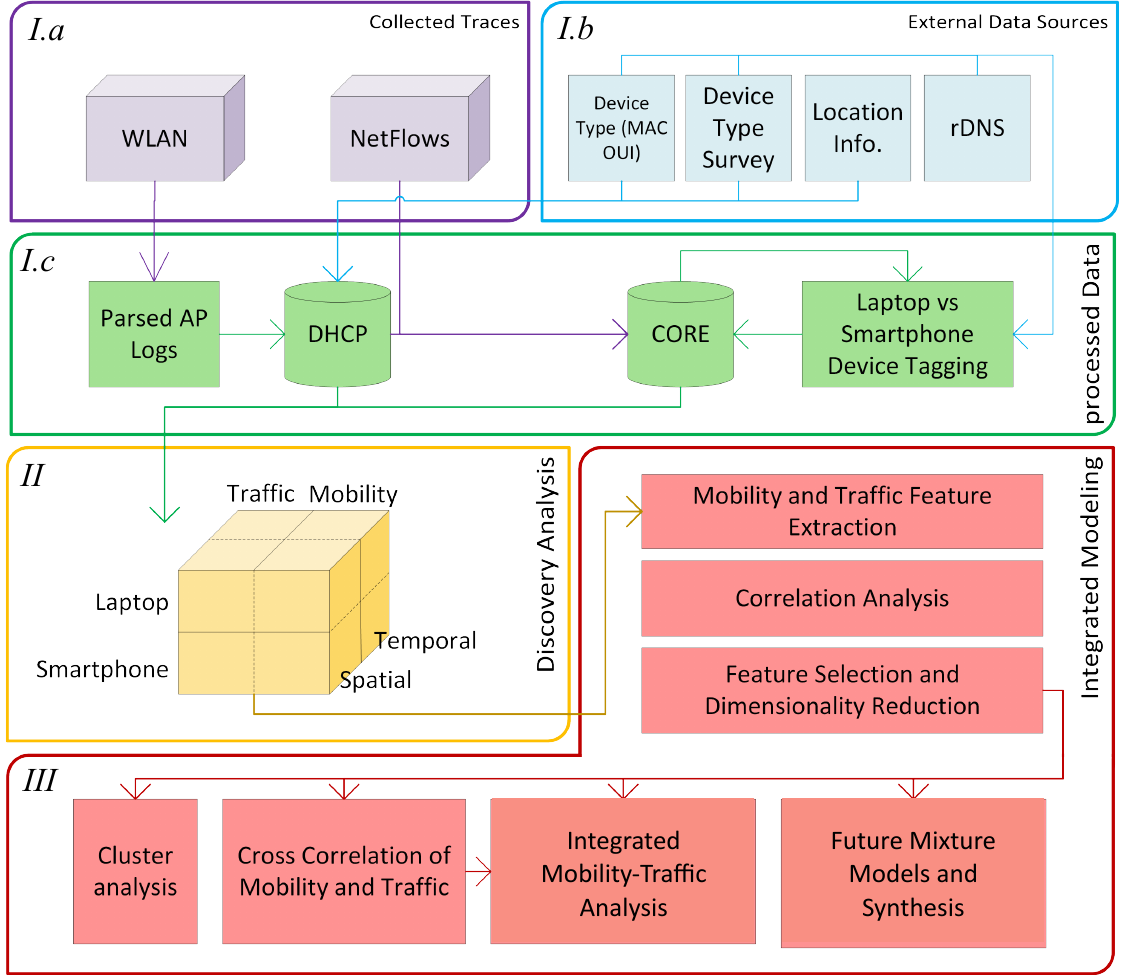}
    \caption{\textit{FLAMeS} system overview.}
    \label{fig:flames}
\end{figure}

To methodically analyze statistical characteristics and correlations in multiple
dimensions, we introduce the \textit{FLAMeS} framework (Fig.  \ref{fig:flames}). 
The main components include: I. Data collection and pre-processing, II. Flutes vs. cellos mobility and traffic analysis, and III. Integrated mobility-traffic analysis.

The two main purposes of this work are to understand and \textbf{quantify} the \textit{gaps between flutes and cellos}, and the \textit{interaction between the mobility and traffic dimensions}. 
Individual mobility and traffic analyses for flutes and cellos are conducted
in Sections V and VI, respectively, with detailed reporting for spatio-temporal
features showing significant gaps.
In Sec. VII, the most important mobility and traffic features are identified
and
 their correlation quantified. 

\section{Experimental setup and datasets}
\label{sec:experiments}

We drive our framework with large-scale datasets from multiple
sources, capturing the mobility and traffic features in different
dimensions.  In this section, we introduce the two data sets and
their preprocessing, and present the device type classification into
flutes and cellos.

%

%


The \textit{input datasets} in this study are specifically chosen to capture: 1. location, mobility
and network traffic information, 2. smartphone and laptop devices, 3. spatio-temporal
features, and 4. scale in number of devices and records. The total size is $>$\textbf{30TB}, consisting of two main parts: WLAN Access Point (AP) logs,
and Netflow records (details in Tables \ref{tab:allnumbers}, \ref{table:netflow_sample}).
\footnote{Data collected using proper procedures. It does not contain \textit{PII}.}

\subsection{WLAN AP logs}
These logs are collected from 1760 APs in 138 buildings over 479 days on a university campus, and contain association and authentication events from 316k devices in 2011-2012. It contains over 555M records, with each record including the device's MAC and assigned IP addresses, the associated AP and a timestamp.
Locations of the APs are approximated by the building locations where they are
installed, i.e., (longitude, latitude) of Google Maps API. To validate this, we fetched
8000 mapped APs around the campus area from a crowd-sourced service, \textit{wigle.net}.
For the 130 matched APs in 42\% of buildings (i.e., 58 bldgs), all were less
than 200m from their mapped location; an error of less than 1.5\% of the campus
area. This is very reasonable for our study purposes.

\subsection{NetFlow logs} \label{sec:netflow}
Over \textbf{76 billion} records of NetFlow traces were collected from the same network, over 25
days in April 2012.  A \textit{flow} is defined as a consecutive
sequence of packets with the same transport protocol,
source/destination IP and port number, as identified by the
collecting gateway router. An example of major Netflow data fields is 
presented in Table \ref{table:netflow_sample}.

The NetFlow records are matched with the wireless associations (from
the AP logs) using the dynamic MAC-to-IP address mapping from the DHCP
logs.  We refer to the result as \textit{CORE} dataset (Table \ref{tab:allnumbers}). They are also augmented with location and website information
using reverse DNS (rDNS)\footnote{Dataset merging and system details in appendix
\ref{sec:appendix_merge} \& \ref{sec:appendix_computing_sys}.}.

\begin{table}[h!]
  \caption{Summary of datasets. B=billion.}
  \begin{subtable}{.70\textwidth}
	\begin{tabular}{c|c|c|c|c|c|c|}
		\cline{2-7}
		& \multicolumn{2}{c|}{\# Records} & \multicolumn{2}{c|}{Traffic Vol. (TB)} & \multicolumn{2}{c|}{\# MAC}  \\ \cline{2-7}
		& DHCP      & CORE    & TCP                & UDP               & WLAN          & CORE                      \\ \hline
		\multicolumn{1}{|c|}{\textit{Flutes}} & 412.0 M   & 2.13 B  & 56.18              & 4.50              & 186.0 K       & 50.3 K                        \\ \hline
		\multicolumn{1}{|c|}{\textit{Cellos}}     & 101.0 M   & 4.20 B  & 73.85              & 12.90             & 93.2 K        & 27.1 K                                                                                                 \\ \hline
		\multicolumn{1}{|c|}{Total}      & 557.5 M   & 6.53 B  & 134.39             & 17.61             & 316.0 K       & 80.0 K                                                                                                 \\ \hline
	\end{tabular}
  \end{subtable}%
  \label{tab:allnumbers}
\end{table}

\subsection{Device type classification}

To classify devices into flutes and cellos, we utilize several
observations and heuristics. To start, note that a device manufacturer (with OUI) can
be identified based on the first 3 octets of the MAC
address\footnote{MAC address randomization does not affect our association trace.}.  Most manufacturers produce one type of device (either
laptop or phone), but some produce both (e.g., Apple).  In the latter
case, OUI used for one device type is not used for another.  We
conducted a survey to help classify 30 MAC prefixes accurately.  Using
OUI and survey information, we identify and label 46\% of the total
devices (90k cellos and 56k flutes).  Then, from the NetFlow logs of
these labeled devices, we observe over 3k devices (92\% of which are
flutes) contacting \textit{admob.com}; an ad platform serving mainly
smartphones and tablets (i.e. flutes). This enables further
classification of the remaining MAC addresses.  Finally, we apply the
following heuristic to the dataset: (1) obtain all OUIs (MAC prefix)
that contacted \textit{admob.com}; (2) if it is unlabeled, mark it as
a flute.  Overall, over 270k devices were labeled (180k as flutes),
covering 86\% of the devices in AP logs and 97\% in NetFlow traces, a reasonable coverage for our purposes.
Out of $\approx80k$ devices in the NetFlow logs, $\approx50$K are flutes and $\approx27$K cellos.

\begin{figure}[ht]
  \centering
  \setlength{\belowcaptionskip}{-10pt}
  \includegraphics[width=0.98\linewidth]{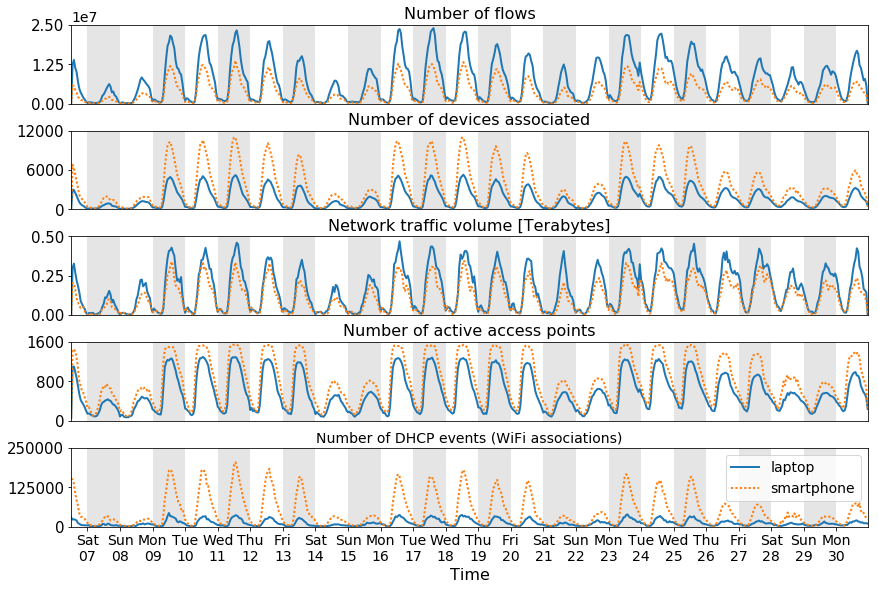}
  \caption{Time series for 25 days of combined AP-NetFlow Core traces}
  \label{fig:timeseries}
\end{figure}

Fig. \ref{fig:timeseries} shows the temporal plot for the combined traces over 25 days, after device classification.
Throughout, the number of flows and total traffic volume is clearly higher for cellos, even with an overall
higher number of flutes connected.
Also note the device activities in a \textit{diurnal} and \textit{weekly} cycles, with the peaks occurring during weekdays, as expected.  Wed, 25th, was the last day of classes, explaining the decline in network activity afterwards.
This plot motivates our analyses for \textit{flutes} vs \textit{cellos}, over \textit{weekends} vs \textit{weekdays}, in this study.

\begin{table*}[h!]
\centering
\caption{NetFlow (top) and AP logs/DHCP (bottom) sample data}
\begin{adjustbox}{max width=\textwidth}
  \begin{tabular}{*{10}{|c}|}
\hline
  Start time & Finish time & Duration & Source IP & Destination IP & Protocol& Source port & Destination port & Packet count & Flow size  \\
\hline
1334332274.912 & 1334332276.576 & 1.664 & 173.194.37.7 & 10.15.225.126 & TCP & 80 & 60482 & 157 & 217708\\
\hline
\end{tabular}
\end{adjustbox}
  \label{table:netflow_sample}
  \vspace*{-0.4cm}
\end{table*}
\begin{table*}[h!]
\centering
\begin{adjustbox}{max width=\textwidth}
  \begin{tabular}{*{6}{|c}|}
\hline
  User IP & User MAC & AP name & AP MAC & Lease begin time & Lease end time  \\
\hline
10.130.90.3 &00:11:22:33:44:55& b422r143-win-1& 00:1d:e5:8f:1b:30 &1333238737&  1333238741\\
\hline
\end{tabular}
\end{adjustbox}
  \label{table:dhcp_sample}
  \vspace*{-0.2cm}
\end{table*}

\section{Mobility analysis}
\label{sec:mobility}

This section covers the \textit{temporal} and \textit{spatial} mobility analyses. For all metrics, unless otherwise noted, we investigate 479 days. A summary of studied metrics and their most significant statistical values are presented in Tab. \ref{tab:mob_numbers} along with mean and median ratios for comparison. From that list, we further investigate in this section those metrics that show the most interesting or non-trivial differences between \textit{flutes} and \textit{cellos}. 

\begin{table}[]
	\centering
	\caption{General results for mobility. Upper values are for weekdays and \red{lower
  ones for weekends} (in red color). \textbf{LJM}: maximum jump [m]; \textbf{DIA}:
  diameter [m]; \textbf{TJM}: total trajectory length [m]; \textbf{GYR}: radius of gyration [m]; \textbf{BLD}: no. uniq. buildings; \textbf{APC}: access point count; \textbf{PDT}: time spent at preferred building [minutes];
	\textbf{DLT}: total session time at each building.}
	\label{tab:mob_numbers}
	\begin{tabular}{|c|c|c|c|c|c|c|c|c|}
		\hline
		\multirow{2}{*}{}             & \multicolumn{3}{c|}{\textbf{Flutes (F)}}                   & \multicolumn{3}{c|}{\textbf{Cellos (C)}}                     & \multicolumn{2}{c|}{\textbf{Ratio  (C/F)}}    \\ \cline{2-9}
		& \textbf{$\mu$} & \textbf{\textit{mdn}} & \textbf{$\sigma$} & \textbf{$\mu$}  & \textbf{\textit{mdn}}  & \textbf{$\sigma$} & \textbf{$\mu$} & \textbf{\textit{mdn}} \\ \hline
		\multirow{2}{*}{\textbf{LJM}} & 435            & 296                   & 813               & 178             & 1                      & 624               & 0.409          & \textbf{0.003}        \\
		& \red{350}      & \red{168}             & \red{683}         & \red{97}        & \red{1}                & \red{312}         & \red{0.277}    & \red{0.006}  \\ \hline
		\multirow{2}{*}{\textbf{DIA}} & 549            & 411                   & 874               & 195             & 1                      & 642               & 0.355          & \textbf{0.002}        \\
		& \red{425}      & \red{179}             & \red{739}         & \red{107}       & \red{1}                & \red{338}         & \red{0.252}    & \red{0.006}  \\ \hline
		\multirow{2}{*}{\textbf{TJM}} & 1582           & 707                   & 2336              & 378             & 1                      & 1444              & 0.239          & \textbf{0.001}        \\
		& \red{1036}     & \red{279}             & \red{1793}        & \red{252}       & \red{1}                & \red{1766}        & \red{0.243}    & \red{0.004}  \\ \hline
		\multirow{2}{*}{\textbf{GYR}} & 396            & 290                   & 2725              & 321             & 191                    & 3265              & 1.102          & 1.019        \\
		& \red{330}      & \red{248}             & \red{1368}        & \red{178}       & \red{65.1}             & \red{1800}        & \red{1.247}    & \red{1.4}    \\ \hline
		\multirow{2}{*}{\textbf{BLD}} & 5.4            & 3                     & 5.6               & 1.8             & 1                      & 2.1               & 0.811          & 0.659        \\
		& \red{2.8}      & \red{2}               & \red{4.1}         & \red{1.5}       & \red{1}                & \red{1.8}         & \red{0.539}    & \red{0.262}  \\ \hline
		\multirow{2}{*}{\textbf{APC}} & 11.8           & 6                     & 13.3              & 3.7             & 2                      & 4.8               & 0.333          & 0.333        \\
		& \red{7.2}      & \red{4}               & \red{8.8}         & \red{3}         & \red{2}                & \red{3.8}         & \red{0.536}    & \red{0.5}    \\ \hline
		\multirow{2}{*}{\textbf{PDT}} & 225            & 161                   & 219               & 248             & 164                    & 254               & 0.314          & 0.333        \\
		& \red{223}      & \red{135}             & \red{272}         & \red{278}       & \red{189}              & \red{292}         & \red{0.417}    & \red{0.5}    \\ \hline
		\multirow{2}{*}{\textbf{DLT}} & 316            & 235                   & 302               & 316             & 217                    & 305               & 1              & 0.92        \\
		& \red{326}      & \red{247}             & \red{308}         & \red{316}       & \red{221}              & \red{309}         & \red{0.97}    & \red{0.89}    \\ \hline
	\end{tabular}
\end{table}

\begin{figure}[ht]
	\centering
	\setlength{\belowcaptionskip}{-14pt}
	\includegraphics[width=0.98\linewidth]{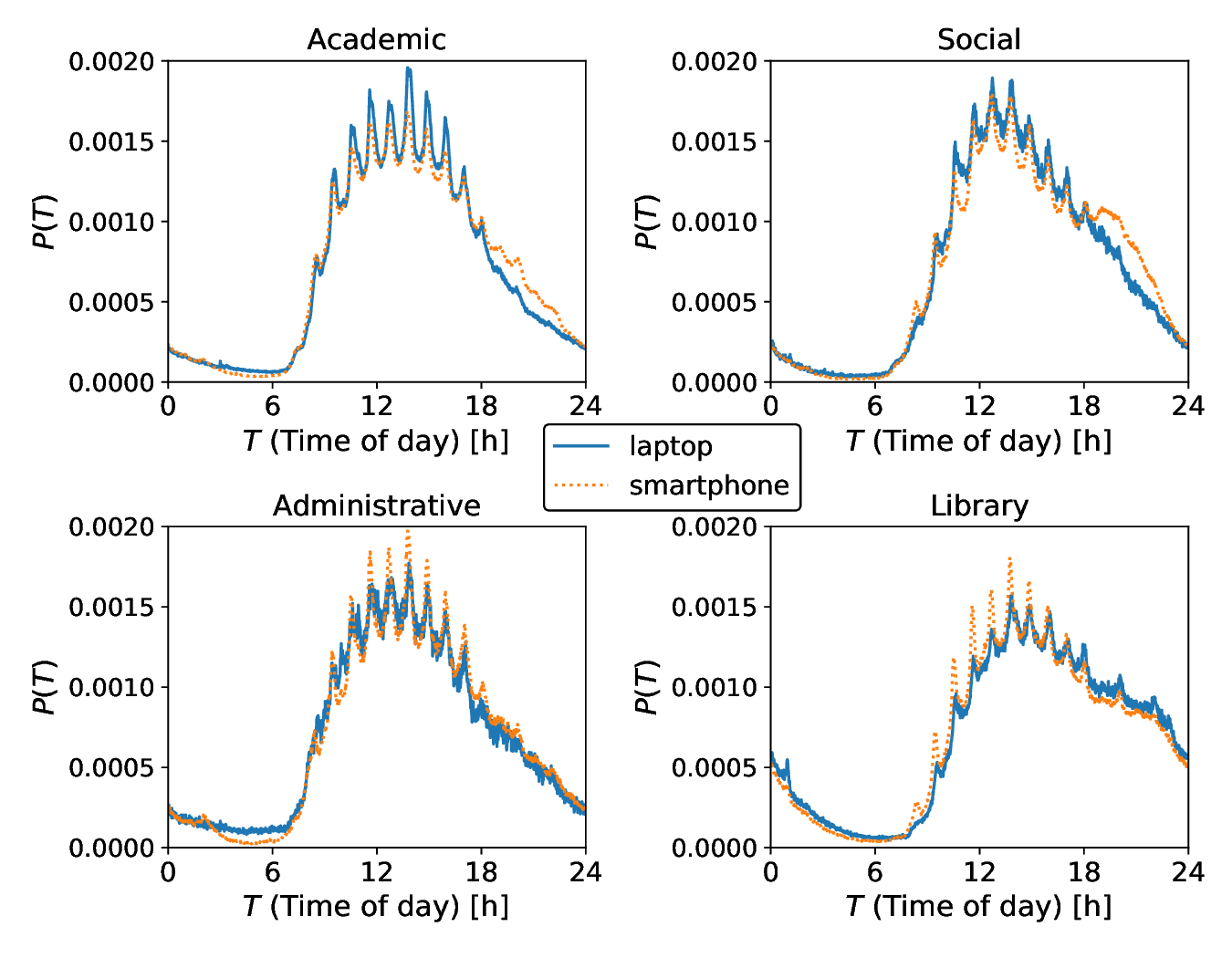}
	\caption{PDF Session start over time of the day.}
	\label{fig:load_building_type}
\end{figure}



\subsection{Session start probability}

A session is defined as the period between WLAN associations.  The
distributions of session start times across the day for four building
categories are depicted in Fig. \ref{fig:load_building_type}.
The start times of the Sessions
match the periodic beginning of classes, but mainly in
\textit{Academic} buildings, where users move mostly at the start and
end of classes.  In these places, activity drops sharply for
\textit{cellos} at 5pm, with considerable \textit{flutes} activity
until 8pm.  For \textit{Social} and \textit{Library} buildings, \textit{the
probability of new sessions remains higher for a few more hours into
the evening, and the times users tend to leave are more spread out}.
We do not make similar observation during weekends, which is expected
when the day is, unlike weekdays, not governed by a class schedule.
For most visitors, the session start distributions show a smooth shape
and no significant differences between device types (omitted for
brevity).
%



\subsection{Radius of gyration}
This metric, $GYR$, captures the size of the geospatial dispersion of
a device's movements,
denoted by $r_g$ and computed as $r_g = \frac{1}{N}
\sum_{k=1}^{N} \left ( \vec{r_k} - \vec{r_s} \right )^{2}$, where
$\vec{r_1}, ..., \vec{r_N}$ are positional vectors of a device and
$\vec{r_s}$ is its center of gravity.

Grouping devices by their $r_g$ after six months of observation, we
look at its evolution since the first time they are observed.
Unsurprisingly (cf. \cite{gonzalez2008understanding}), after an
initial transient period of about one week, this value stabilizes even
across different semesters (not shown).  

\begin{figure}
	\centering
	 \setlength{\belowcaptionskip}{-14pt}
	\captionsetup[subfigure]{aboveskip=-2.5pt,belowskip=-2.5pt}
	\begin{subfigure}[b]{0.49\linewidth}
		\includegraphics[width=\textwidth]{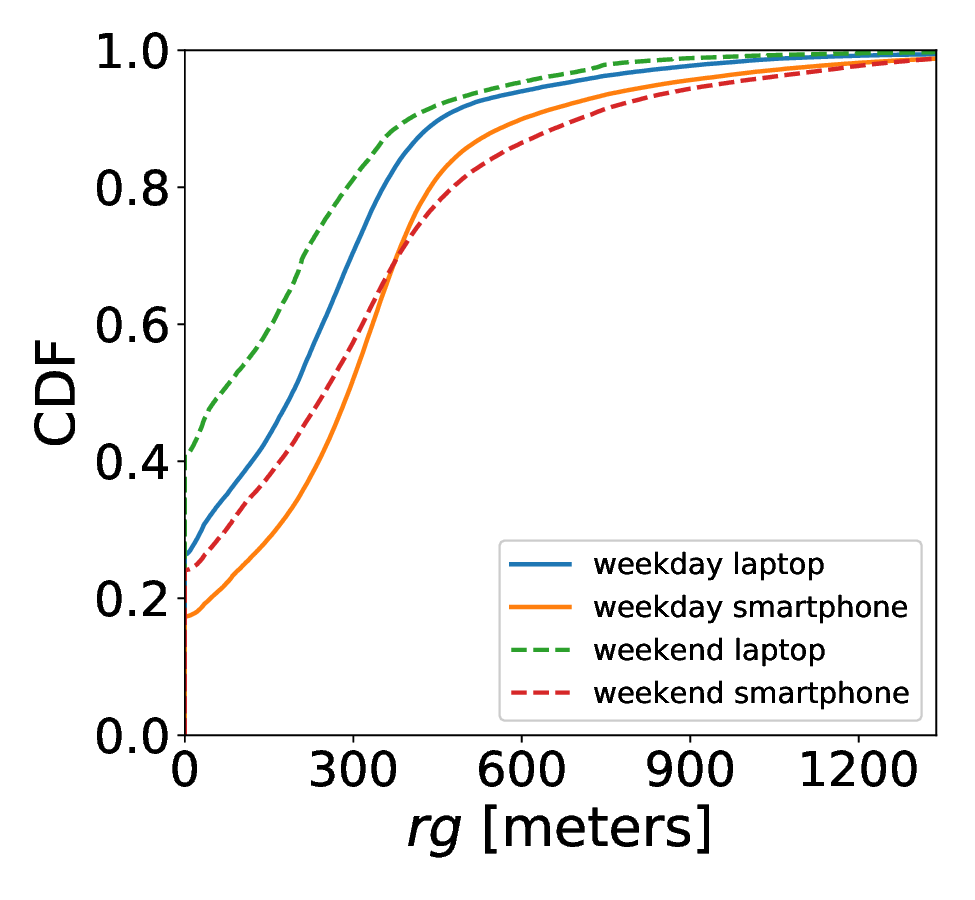}
		\caption{}
		\label{fig:rgweek}
	\end{subfigure}
	\begin{subfigure}[b]{0.49\linewidth}
		\includegraphics[width=\textwidth]{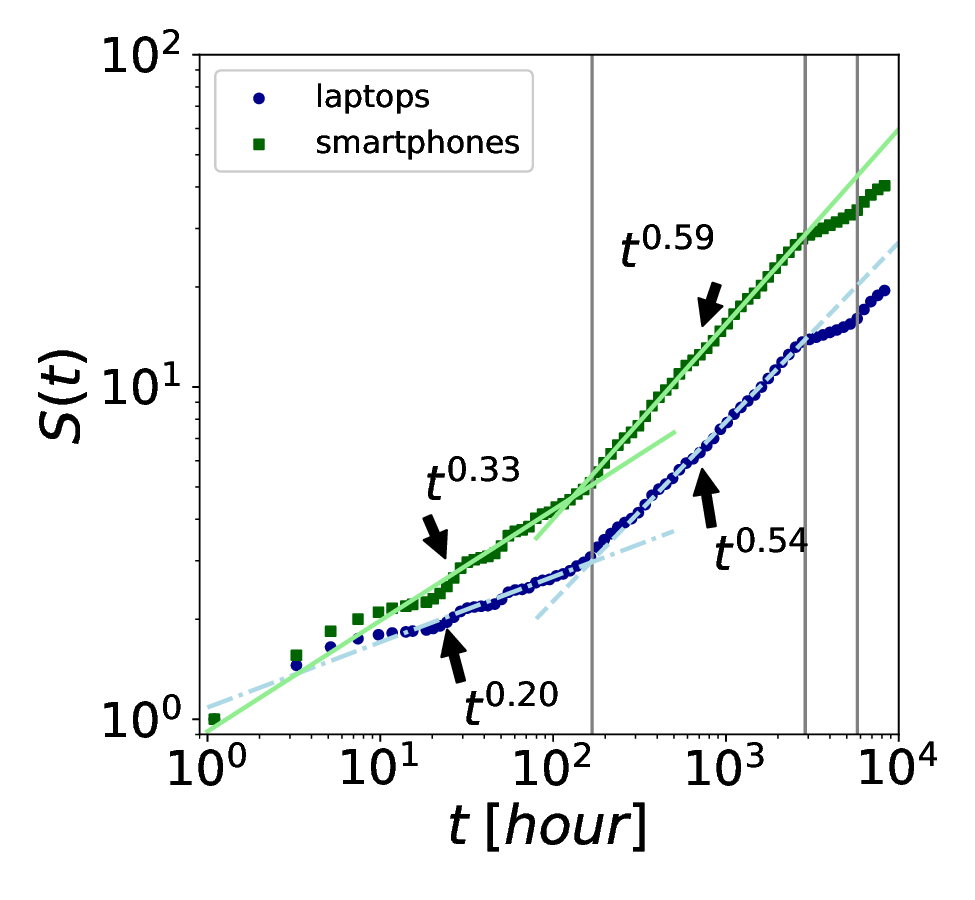}
		\caption{}
		\label{fig:visitedlocs}
	\end{subfigure}
	\caption{(a) Radius of gyration ($rg$ for the device types). 
		(b) Visited locations $S\left(t\right)$. Vertical lines at 7, 120 and 240 days.}
	\label{fig:rg_st}
\end{figure}

We split the traces into weekdays and weekends, presenting the
distributions in Fig. \ref{fig:rgweek}. 
For \textit{cellos}, we notice a substantial reduction in their overall
mobility whereas, for \textit{flutes}, this difference is not so
pronounced.  This might be due to students having fewer activities on
weekends, a tendency to study at a single building like a library, or
just not carry their cellos; we will revisit this aspect in Sec.
\ref{sec:multi_dim}.  \textit{Flutes}, being ``always-on'' devices,
are able to capture movements at pass-by locations, dining areas, and
bus stops and thus are better suited to capture the fine-granular
mobility of their users than cellos.

Despite the 8.1km$^2$ area of the campus (approximate radius of 1.42km), 
buildings with related fields of study (e.g. Fine Arts) are fairly
clustered.
Computing the distance between the k-nearest neighboring buildings,
for $k=22$ and $k=9$ (average number of visited buildings for
\textit{flutes} and \textit{cellos})
the median distances are 295m and 172m, respectively.  
Due to their focus on classes, attending students have limited area of activity on
weekdays, which explains the observed \textit{radius of gyration}.

We also evaluated: (1) \textit{diameter} $DIA$, the longest distance
between any pair of $\vec{r_k}$ points; (2) \textit{max jump} $LJM$,
the longest distance between a pair of consecutive $\vec{r_k}$ points;
and (3) \textit{total trajectory length} $TJM$, the sum of all trips
made by a device.  The distributions of these metrics are similar to
\textit{Radius of Gyration} and therefore not shown.  Table
\ref{tab:mob_numbers} summarizes the most significant statistical
values for these metrics.

\subsection{Visitation preferences and interests}

We count the number of unique buildings visited by a user, $BLD$, and
define a \textit{preferred building} as the location where a device
has spent most of its time in a given day, measured in minutes
and referred to as $PDT$.  We approximate the latter by the formula $t_b
= \sum_{k=1}^{N_b} S_k$, where $t_b$ is the time spent, $N_b$ the
total number of sessions and $S_1 ... S_N$ the time duration of \textit{each
session} at a building $b$, here referred as $DLT$. Interestingly, 
cellos have slightly longer stays but both have medians around 2:40 hours.
The similarity of the distributions, combined with a lower number of visited
locations indicate that cellos are used mostly when users remain longer periods
at places.

Fig. \ref{fig:visitedlocs} highlights the differences between
\textit{flutes} and \textit{cellos} on the required time $t$ to visit
\textit{S}(\textit{t}) locations. \textit{After an initial exploration period
of one week the rates of new visits change similarly for both device
types, and new exploration rates show up at 120 and 240 days.} These
could be explained by the weekly schedules of the university as well
as the usual length of a lecture term ($\approx 4$ months).


\begin{figure}[ht!]
	\centering
	\setlength{\belowcaptionskip}{-10pt}
	\includegraphics[width=0.47\textwidth]{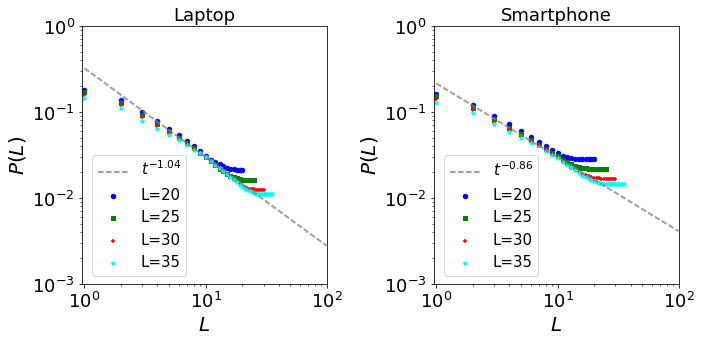}
	\caption{Zipf's plot on $L$ visited access points.}
	\label{fig:pl_visited}
\end{figure}

We also consider the number of unique APs a device
associates with, $APC$, which provides a finer spatial resolution than
the building level. Furthermore,  the probability of finding a device
at its \textit{L-th} most visited access point is shown in
Fig. \ref{fig:pl_visited}. When taking buildings as aggregating points
for location, the values become $L^{-1.36}$ for \textit{cellos} and
$L^{-1.16}$ for \textit{flutes}. These approximations validate
previous work on human mobility \cite{gonzalez2008understanding}, yet 
highlight differences between device types.

\subsection{Sessions per building}

\begin{figure}[ht!]
	\centering
	\setlength{\belowcaptionskip}{-15pt}
	\includegraphics[width=0.47\textwidth]{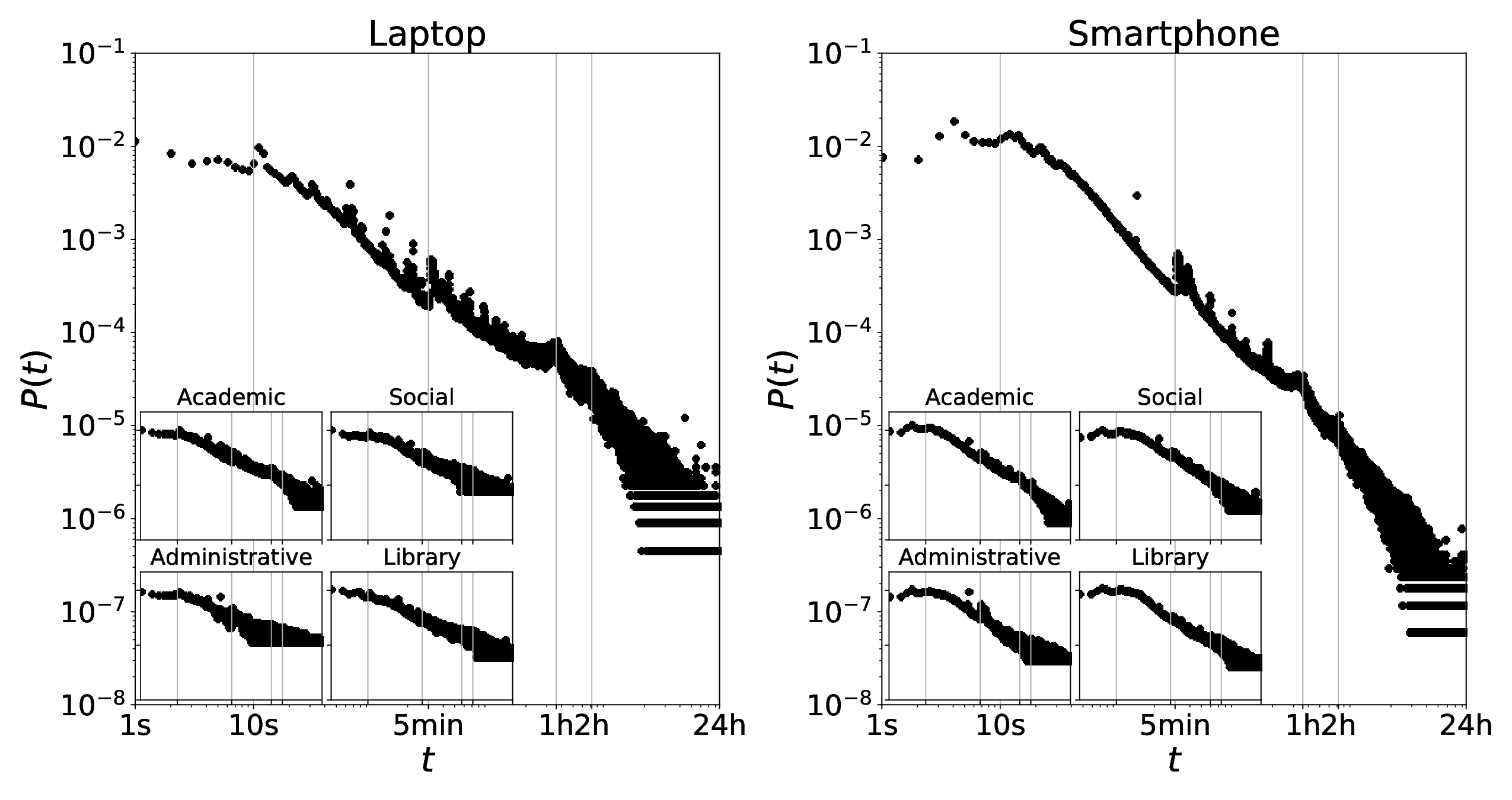}
	\caption{Probability $P\left(t\right)$ of session duration $t$.} 
	\label{fig:pt_ap}
\end{figure}

To study AP utilization over time, we look at the session duration
distribution, or session duration dispersal kernel P(t), depicted in
Fig. \ref{fig:pt_ap}.
The smaller inner plots represent the same metric, limited to four
types of buildings. 

We noted that the five-minute spikes correspond to default
idle-timeout for the used WiFi routers.  On the other hand, the
\textit{knees} at 1 and 2 hours could be explained by the typical
duration of classes.  They are only noticeable at Academic buildings
(shown inside inner plots) and during weekdays (not shown).  This
leads us to conclude that despite the differences in distributions of
device types, \textit{flutes} and \textit{cellos} present \textit{certain
similarities in their usage, such as during classes}.  To differentiate
\textit{pass-by} access points, we examine all sequences of three
unique APs where all session durations are lower than 5 minutes
(typical idle-timeout). We observed these APs clustered at buildings
that also had major bus stops nearby.  


\section{Traffic analysis}
\label{sec:traffic_analysis}

In this section, we compare different \textit{traffic} characteristics, across \textit{device types},
\textit{time} and \textit{space}.
For this purpose, we start  with statistical characterization of \textit{individual} flute and cello flows.
Next, we measure how these flows, \textit{put together}, affect the network patterns across APs and buildings.
Finally, \textit{user behavior} is analyzed by monitoring weekly cycles, 
data rates, and active durations. 
By quantifying \textit{temporal} and \textit{spatial} variations of traffic across device types,
we make a case for new models to capture such variations based on the most relevant attributes.
Table \ref{tab:traff_numbers} summarizes the results.


\begin{figure}[!ht]
	\begin{subfigure}{0.495\linewidth}
		\includegraphics[width=\linewidth]{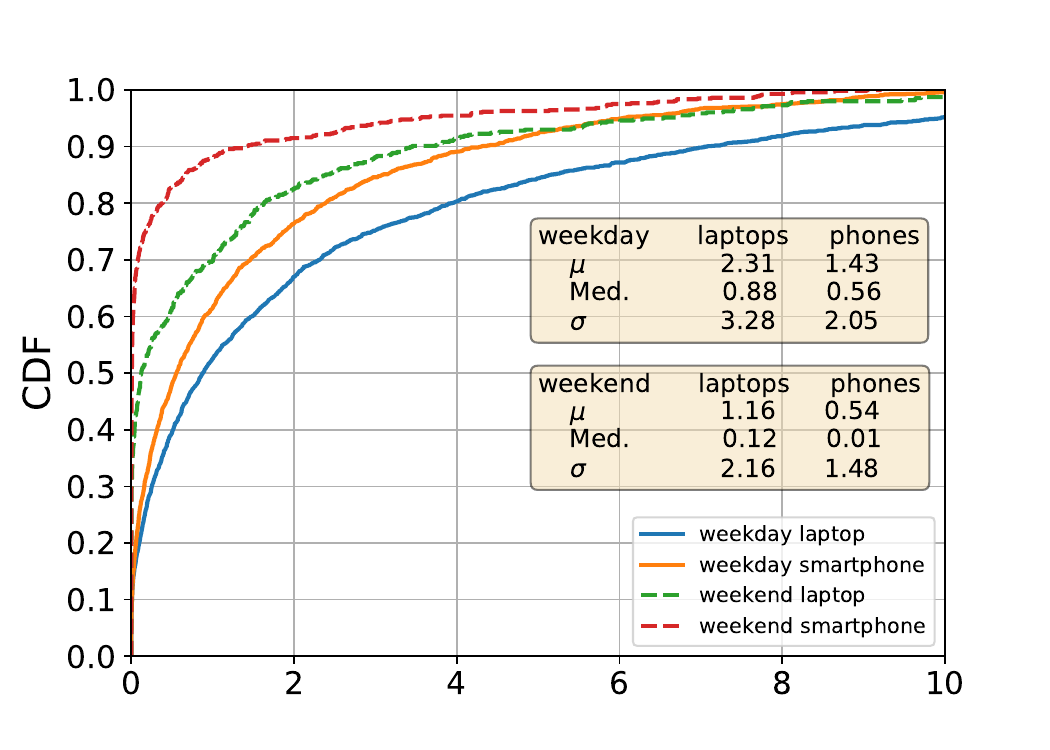}
		\caption{Packet processing rate of APs \\(millions per day)}
		\label{fig:aps_weekDAYANDEND_packets_CDF}
	\end{subfigure}	
	\begin{subfigure}{0.495\linewidth}
		\includegraphics[width=\linewidth]{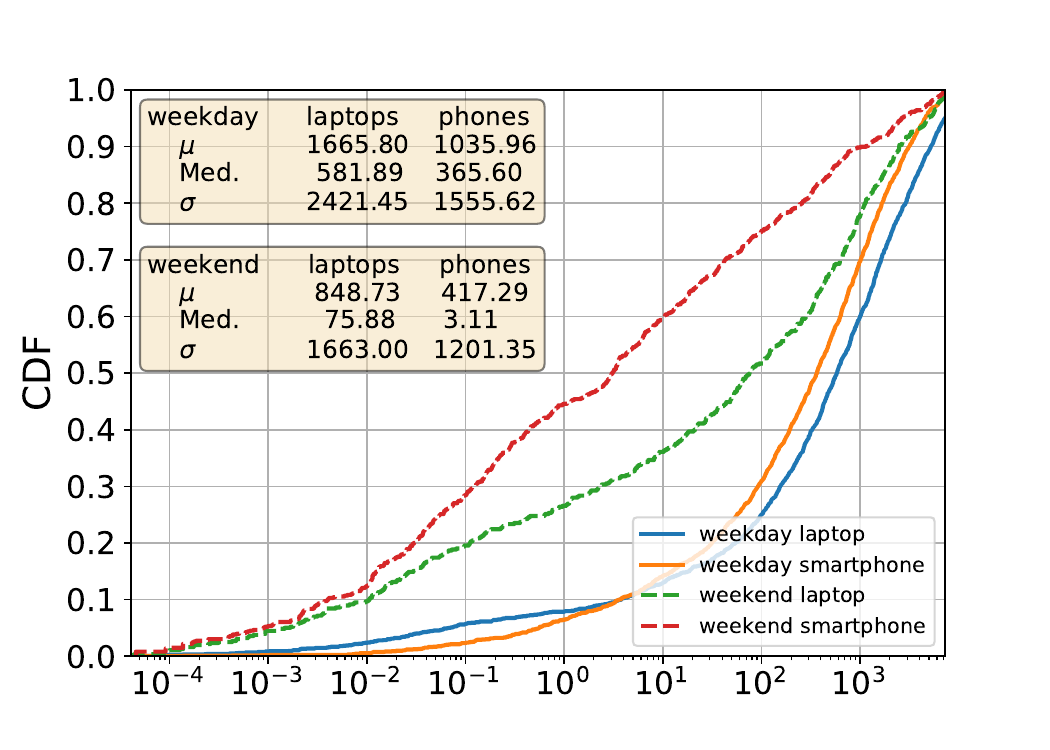}
		\caption{Traffic load of APs \\(MB per day, log-scale)}
		\label{fig:aps_weekDAYANDEND_bytes_CDF}
	\end{subfigure}     
	
		\begin{subfigure}{0.495\linewidth}
		\setlength{\abovecaptionskip}{-5pt}
		\setlength{\belowcaptionskip}{5pt}
		\includegraphics[width=\linewidth]{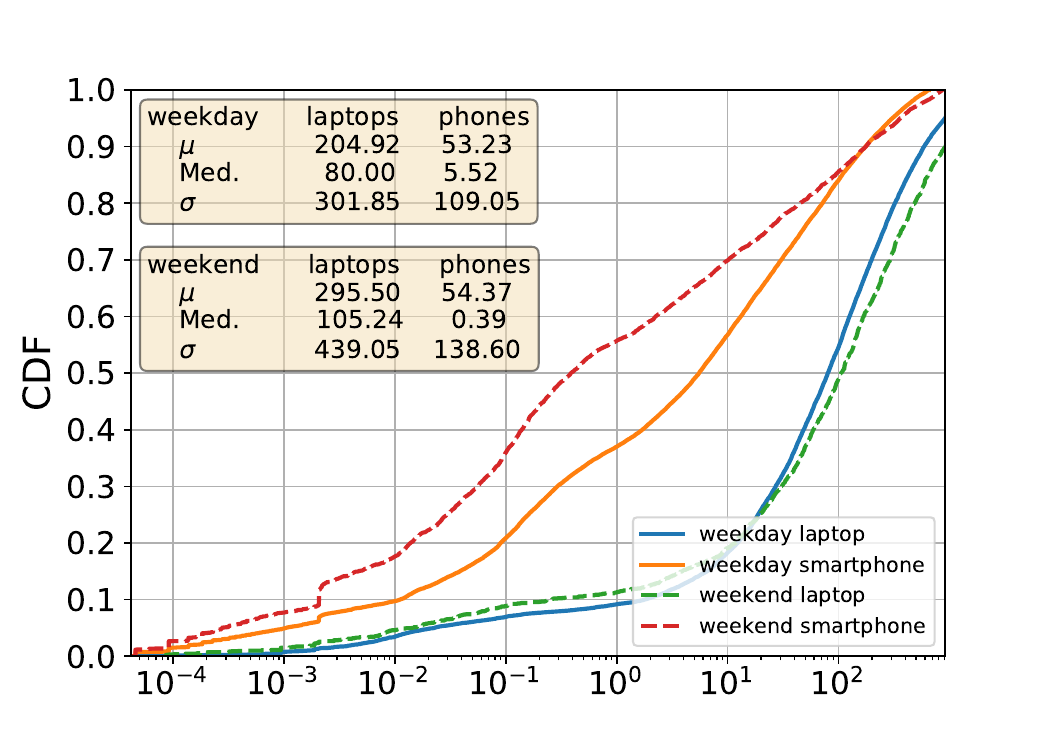}
		\caption{User data consumption \\(MB per day, log-scale)}
		\label{fig:users_weekDAYANDEND_bytes_CDF}
	\end{subfigure}
	\begin{subfigure}{0.495\linewidth}
		\setlength{\abovecaptionskip}{-5pt}
		\setlength{\belowcaptionskip}{5pt}
		\includegraphics[width=\linewidth]{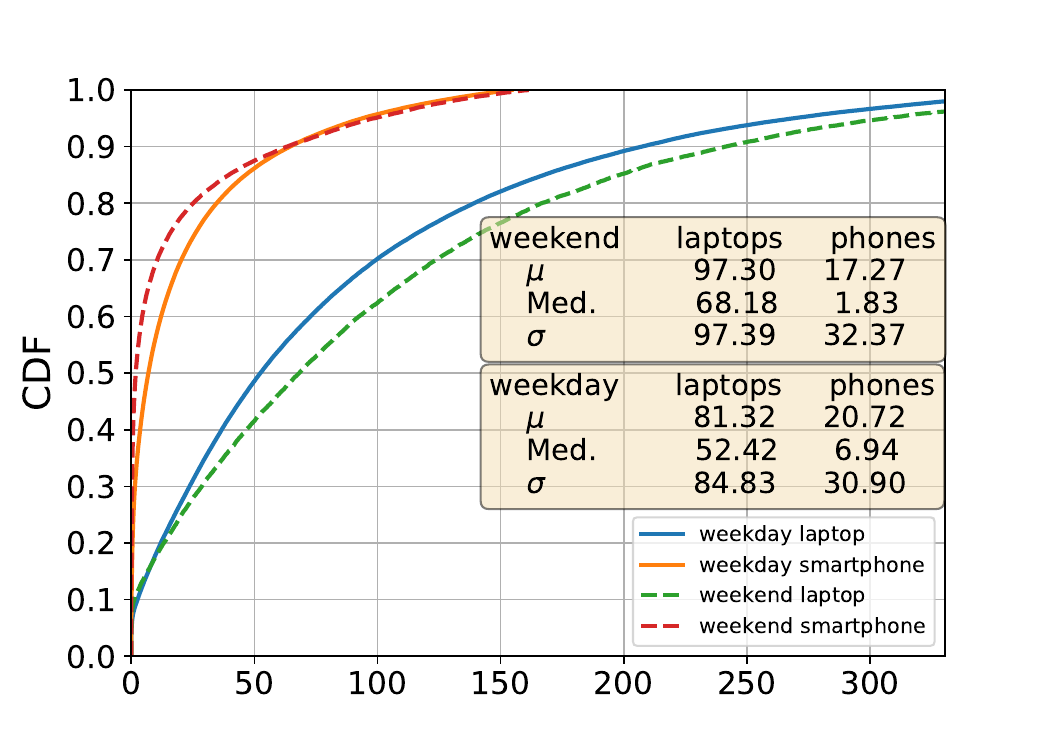}
		\caption{User active time \\(minutes per day)}
		\label{fig:users_weekdays_active_time_CDF}
	\end{subfigure} 
	\caption{Distribution plots}
\end{figure}

\subsection{Flow-level statistical characterization}
We compare the following distributions using maximum likelihood estimation (MLE) 
and maximizing goodness-of-fit estimation: Gaussian, Exponential, Gamma, Weibull, Logistic, Beta and Lognormal\footnote{For distribution comparison, significance threshold $p-value$ is set at $.05$.}.

\subsubsection{Size}
Flow size is the sum of bytes for all packets within a single flow.
On weekdays, average \textit{size of individual flute flows} is \textbf{$>2x$ larger} \textit{than cello flows} ($2070$ vs. $822$ bytes),
while median is \textbf{$>4x$ larger} ($678$ vs. $142$ bytes).
There are no significant changes on weekends.

The average packet size within a single flow also provides insight into packet-level
behavior of services on mobile devices. 
We notice that the average \textit{packet size of flute flows} is \textbf{$\approx$50\% larger} \textit{than that of cellos}
(212 vs 144 bytes on weekdays, 205 vs 142 on weekends).
Comparing weekdays and weekends, median size of flute packets drops on weekends whereas it remains \textit{the same} for cellos.
In fact, comparing cello flows on weekdays and weekends shows \textit{no significant difference} in terms of average packet size (p-value$>.05$).
Despite smaller flows, the average cello generates \textit{2.7 times traffic} as an average 
flute because the average cello is responsible for \textit{3.7 as many flows} as a flute.
Analyzing distributions of flow size and average packet size  in our datasets shows 
that \textit{Lognormal} distribution is the best fit,  with varying parameters for each device type  (More details in Sec. \ref{sec:appendix_traffic} of appendices). 

\subsubsection{Packets}
This metric is the count of packets within each flow. The mean and median packet counts per flow are $7.06$ and $5$ in flutes and $3.64$ and $2$ in cellos, during weekdays. The means drop slightly on weekends. Packet counts per flow match the \textit{Lognormal} distribution well for flows of both device types.
The average flute flow is bigger in size and has \textit{more packets} (with higher variance) but there are \textit{fewer} flows coming from these devices.
This is analyzed further for TCP/UDP flows (Sec. \ref{sec:protocols}).
   
\subsubsection{Runtime}
Flow runtime is the period of time the flow was active (equal to a flow's $finish time -  start time$).  
Flute flows have a mean and median of $1868$ms  and $128$ms respectively on weekdays, while these numbers are $1639$ms and $64$ms for cellos. 
Both device types show increase in means during weekends (flutes by $204$ and cellos by $164$), 
indicating that although there are fewer devices online during weekends, they are more active.
The low medians in either group corresponds to many \textit{short-lived} flows with few packets,
showing little variation across device type, time or space.

\subsubsection{Inter-arrival times (IAT)}

Median of the flow \textit{IAT}\footnote{IAT is important in simulation and modeling of networking protocols, traffic classification \cite{Moore2005}, congestion control and traffic performance \cite{Paxson1995}. Our flow-level IAT analysis can also be used for measuring delay and jitter effects.} at APs is $6$ms for cello flows and $4$ms in case of flutes, on weekdays (similar on weekends), 
which suggests that the majority of APs handle flows from either device type at nearly the same rate.
However, average \textit{IAT} is $\approx143$ms for flute and $\approx78$ms for cello flows, as there are more cellos with very high rate of flows.
Flow \textit{IAT} in our datasets matches a \textbf{beta} distribution well (See appendix \ref{sec:appendix_traffic}) 
with a \textit{very high estimated kurtosis} and \textit{skewness} (estimated at 58 \& 6.9 respectively).
The high estimated kurtosis illustrates that there are \textit{infrequent extreme values}, which explains the observed highly elevated standard deviation of \textit{IAT}.
Higher average IAT of flutes, combined with the higher standard deviation compared to cellos (596 vs 284), 
shows that \textit{flutes face more extreme periods of inactivity, which can be caused by higher mobility and packet loss}.

\subsubsection{Protocols}
\label{sec:protocols}
TCP accounts for \textit{78.5\%} of cello flows (\textbf{84.6\% of bytes}) and \textit{98.2\%} of flute flows (\textbf{91.6\% of bytes}). 
The higher presence of UDP in cellos is reasonable, considering that UDP applications (e.g., multi-player games, video conferencing and file sharing) are more likely to be used with cellos.
Comparing the number of packets in flows, in case of TCP, the average number of packets in cello flows is almost half that of a flute flow (4.6 vs 8.8), and the average packet size of flutes is 22\% higher than that of cellos.
This supports our earlier observation regarding the bigger flow sizes of flutes.
However, for UDP, the two device types are similar in terms of
average packet count per flow (2.5 for cellos \& 2.87 for flutes) and average packet size (119 for both).
This conforms to low latency requirements of many UDP applications.
\par
Given these differences, traffic classification using machine learning \cite{Nguyen2008} 
could benefit from considering device types to train models. We investigate this in Sec. \ref{modeling}.

After establishing the similarities and differences of flows, the next step 
is to evaluate whether the individual variations in flows lead to different \textit{aggregate traffic behaviors} from 
viewpoint of the network.

\subsection{Network-centric (spatial) analysis}
We examined the load of APs in all buildings on a daily basis to provide insight into differences from the viewpoint of the network.
For each AP, we calculate flow metrics for every weekday and weekend.
We focus our analysis on the first three weeks of NetFlow traces to avoid
significant user behavior change during exams period, as already shown in Fig. \ref{fig:timeseries}.

First, we measure the daily packet and flow arrival rates at APs. 
The median flow rates are $42k$ and $20k$ per weekday for cellos and flutes respectively (7.5k and 0.5k on weekends). 
The average number of cello packets processed daily by APs is \textbf{1.6 times higher} than flute packets (Fig. \ref{fig:aps_weekDAYANDEND_packets_CDF}).
Each AP handles, on average during weekdays, \textit{$\approx27$ cello packets per second} and  \textit{$\approx17$ flute packets per second},
dropping to $\approx13.5$ and $\approx6.25$ on weekends.
This indicates that,  during the weekends, a high percentage of access points are not utilized, 
with \textit{60\% of APs seeing no flute flows} and \textit{70\% receiving no cello flows}. However, \textit{at least one AP in $>$80\% of buildings sees traffic}, supporting observations
of less mobility during weekends.

Next, we look at traffic volume. 
On average weekdays, 90\% of APs handle < $5GB$ of cello traffic ($2.5GB$ on weekends), 
whereas the same percentage handles < $3GB$ of flute traffic ($1GB$ on weekends) (Fig. \ref{fig:aps_weekDAYANDEND_bytes_CDF}). 
Flutes are more mobile, visit a higher number of unique APs and have bigger flow sizes but they are
still responsible for \textit{less overall network load}.

Thus, the individual differences of flute and cello flows result in 
\textit{heterogeneous aggregate traffic patterns} in time (different days) and space (APs at different buildings)\footnote{A more in-depth analysis is presented in appendix \ref{sec:appendix_traffic}.}.
With that established, in order to take steps towards modeling and simulation, we also need to analyze the behavior of users.

\subsection{User behavior (temporal) analysis}
Here, we measure traffic patterns from a user-centric perspective. 
We identified gaps in diurnal and weekly cycles (Fig. \ref{fig:timeseries}) as well as
traffic flow features of individual \textit{users} including data consumption, packet rates, and network activity duration.


\subsubsection{Data consumption}
Fig. \ref{fig:users_weekDAYANDEND_bytes_CDF} shows daily data consumption, with 90\% of cellos consuming \textbf{< 700MB} and 90\% of flutes using \textbf{< 200M} on weekdays.
Surprisingly, for cellos on campus during weekends, average data consumption is even higher whereas data consumption of flutes drops sharply. 
  
\subsubsection{Packet rate}
On weekdays, cellos on average generate \textbf{$\approx$318K packets}, while flutes only average \textbf{ $\approx$84K packets} per day. 
On weekends, the few on-campus cellos see greatly increased number of packets, with an average daily packet rate of $\approx495$K. Weekend flutes also have a modestly \textit{increased} packet count, with an average of $\approx96$K flows.

\subsubsection{Active duration}
Total active time of devices serves well to demonstrate the differences between time spent online by users of different device types.
We rely on NetFlow to measure \textbf{'active'} time instead of AP association time. 
This allows us to distinguish user's \textit{idle} presence in the network from its \textit{activity} periods.
Cellos have \textbf{4x} average active time compared to flutes in our traces ($\approx81$ vs $\approx21$ min on weekdays, $\approx97$ vs $\approx17$ min on weekends). Overall, 90\% of cellos are active for \textbf{$<$3.5h} and 90\% of flutes are active for \textbf{$<$1h} (Fig. \ref{fig:users_weekdays_active_time_CDF}). 
As evident in various metrics, the cellos appearing on weekends are more active than the average cello on weekdays. 

Overall,  the data consumption of flutes seems to be \textit{more bursty} in nature, with \textbf{bigger} flows and \textbf{lower active duration}. 
This could be due to more intermittent usage of flutes and also bundling of network requests to save battery on these devices.
In addition, there are fewer devices on campus during weekends, but those remaining devices are more active and consume more data than average.

\section{Integrated mobility-traffic analysis}
\label{sec:multi_dim}
We study the relation between mobility and network traffic features,
examine whether their \textit{fusion} provides a case for
the necessity of \textit{integrated mobility-traffic models}, and 
introduce steps towards such models (Sec. \ref{modeling}).

\subsection{Feature engineering} \label{sec:feature_engineering}
To simplify analysis and interpretation, and reduce dimensionality, we identify the most important features.
First, we study the relationships among variables from \textit{mobility} and \textit{traffic} dimensions separately.
Then, from this subset of combined features, we investigate whether clusters of user devices appear in the dataset.
For this, we use correlation feature selection (\textit{CFS} \cite{hall1999correlation}), to obtain uncorrelated 
features,
but highly correlated to the classification.
Finally, we quantify correlations between mobility and traffic metrics (See abbreviations in Fig. \ref{fig:correlations}).
Pearson correlation is shown in the figures.

\subsubsection{Mobility}
\begin{figure*}
	\centering
	\small
	\setlength\tabcolsep{4pt}
	\begin{tabular}{cccc}
		\includegraphics[width=0.25\textwidth]{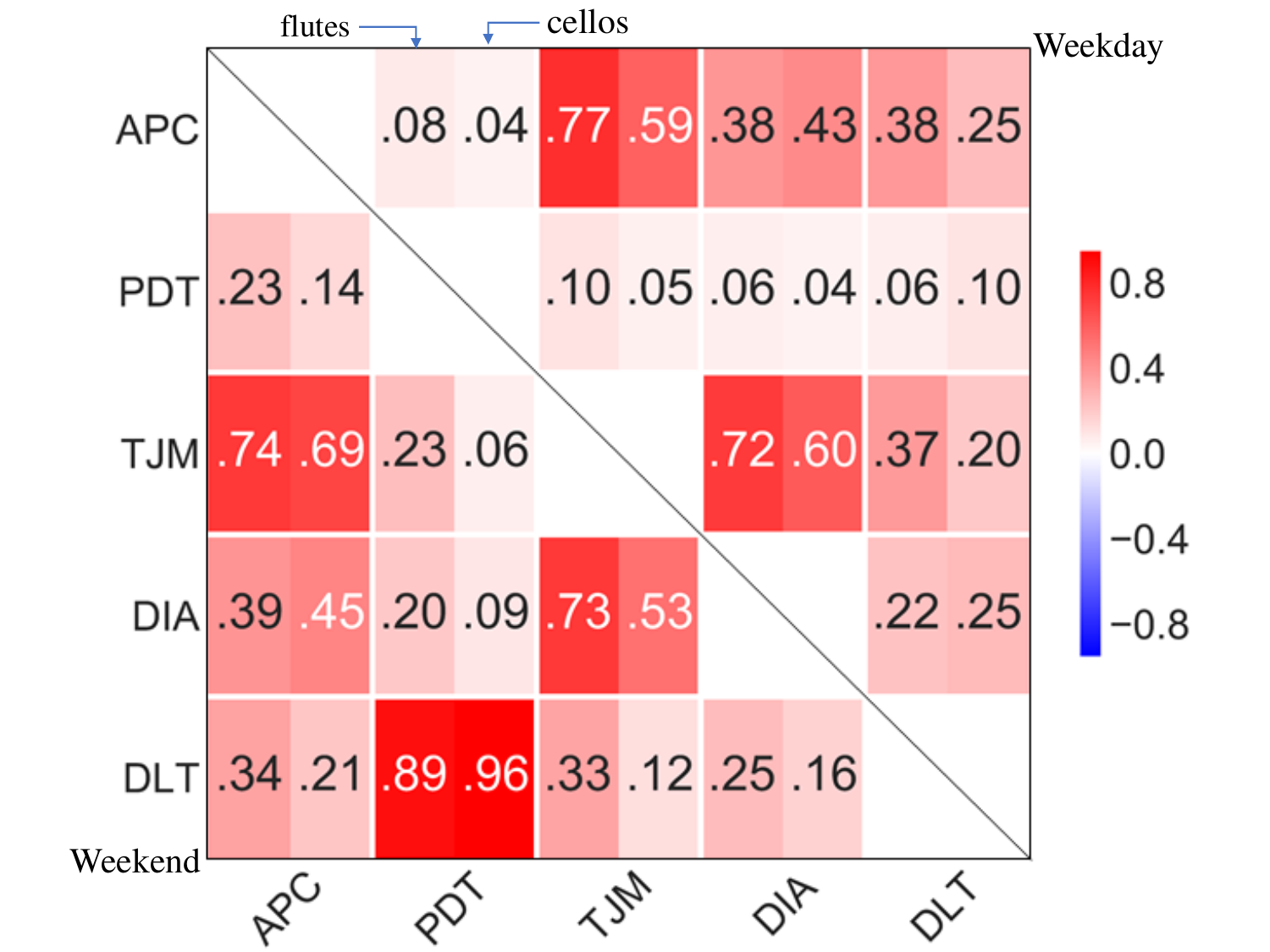}  & & \multirow{2}{*}[1.45in]{\includegraphics[width=0.41\textwidth]{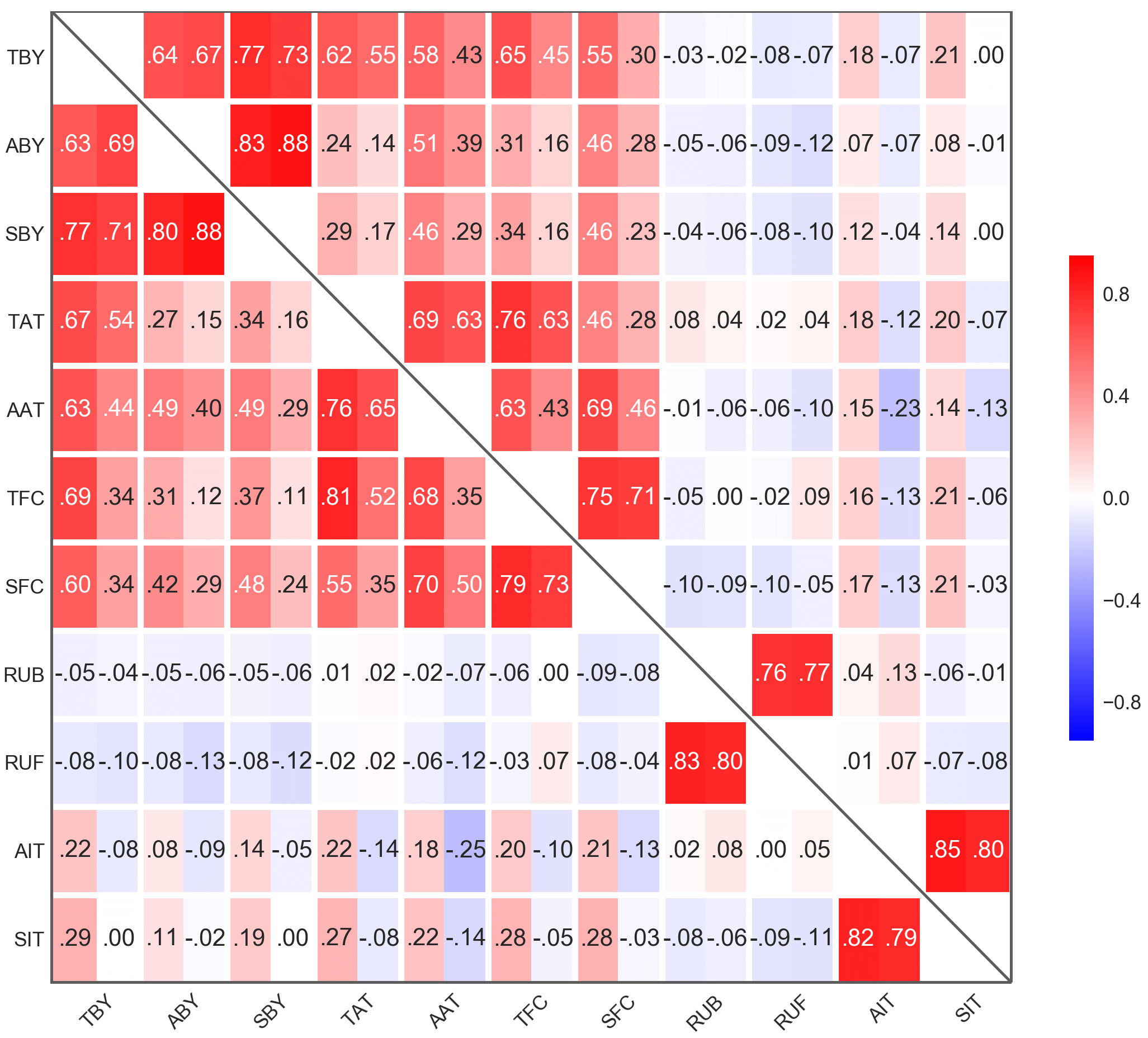}} &
		\multirow{2}{*}[1.3in]{
			\setlength\tabcolsep{3pt}
			\begin{tabular}{|c|p{2.9cm}|}
				\hline
				Abbr. & Description \\
				\hline\hline
				TBY & Total flow bytes                                         \\ \hline
				ABY & Avg. flow bytes                                         \\ \hline
				SBY & Std. flow bytes                                            \\ \hline
				TAT & Total active time                                                           \\ \hline
				AAT & Avg. active time \\ \hline
				TFC & Total flow count                                \\ \hline
				SFC & Std. flow counts                               \\ \hline
				RUB & UDP bytes / total bytes \\ \hline
				RUF & UDP flows / total flows \\ \hline
				AIT & Avg. IAT \\ \hline
				SIT & Std. IAT \\ \hline
		\end{tabular}} \\
		\setlength\tabcolsep{3.2pt}
		\begin{tabular}[0.2in]{|c|p{2.7cm}|}
			\hline
			Abbr. & Description \\
			\hline\hline
			APC & AP Count (unique)                                              \\ \hline
			PDT & Preferred building $\Delta t$                                                 \\ \hline
			TJM & Total (sum) jumps \\ \hline
			DIA & Diameter of mobility                                \\ \hline
			DLT & Delta time (time of network association) \\ \hline
		\end{tabular} &          &                &                 \\
		(a) Mobility & & \multicolumn{2}{c}{(b) Traffic}
	\end{tabular}
	\caption{Correlation plots for (a) \textit{mobility} and (b) \textit{traffic} features. Each cell's left half is for flutes and right half is for cellos, the upper right triangle is for weekdays and the lower left for weekends. \label{fig:correlations}}
\end{figure*}

The \textit{CFS} algorithm was run on 8 features (in Sec. V), and kept only \textit{5} (to be used in the cross-dimension analysis).
Fig. \ref{fig:correlations}a visualizes the linear dependence between mobility features, comparing flutes and cellos on weekdays and weekends.
Close inspection reveals temporal correlation relationships. For example, for cellos on weekends, there is a \textbf{strong} correlation ($0.96$) between preferred building time (\textit{PDT}) and time of network association (\textit{DLT}), but weak correlation ($0.1$) on weekdays, suggesting that most of weekend online time is spent at preferred buildings (e.g., libraries).

\begin{table}[]
\centering
\caption{Traffic features used for integrated mobility-traffic analysis  (per device, per day; 
see Fig. \ref{fig:correlations} for abbreviations).
Upper values are for weekdays and \red{lower ones for weekends} (in red).}
\label{tab:traff_numbers}
\setlength\tabcolsep{1pt}
\begin{tabular}{c|c|c|c|c|c|c|c|c|c|}
\cline{2-10}
                                                                                              & \multicolumn{3}{c|}{Flutes (F)} & \multicolumn{3}{c|}{Cellos (C)}   & \multicolumn{3}{c|}{Ratio (C/F)} \\ \cline{2-10} 
                                                                                              & \textbf{$\mu$}     & \textbf{\textit{mdn}}   & \textbf{$\sigma$}       & \textbf{$\mu$}      & \textbf{\textit{mdn}}    & \textbf{$\sigma$}       & \textbf{$\mu$}     & \textbf{\textit{mdn}}     \\ \hline
\multicolumn{1}{|c|}{\multirow{2}{*}{\begin{tabular}[c]{@{}c@{}}TBY\\ {[}MB{]}\end{tabular}}} & 96.77    & 11.47    & 194.52    & 373.08    & 144.68    & 554.54    & 3.85     & \textbf{12.61}     \\
\multicolumn{1}{|c|}{}                                                                        & \red{80.96} & \red{0.86} & \red{195.15} & \red{448.87} & \red{180.23} & \red{623.86} & \red{5.54} & \red{209.56} \\ \hline
\multicolumn{1}{|c|}{\multirow{2}{*}{ABY}}                                                    & 5.48     & 0.74     & 14.02     & 15.67     & 7.34      & 25.81     & 2.85     & \textbf{9.91}           \\
\multicolumn{1}{|c|}{}                                                                        & \red{4.54} & \red{0.15} & \red{14.16} & \red{18.06} & \red{8.34} & \red{28.71} & \red{3.97} & \red{55.6}  \\ \hline
\multicolumn{1}{|c|}{\multirow{2}{*}{SBY}}                                                    & 10.56    & 1.57     & 23.76     & 30.59     & 13.77     & 49.82     & 2.89     & \textbf{8.77}          \\
\multicolumn{1}{|c|}{}                                                                        & \red{8.09} & \red{0.13} & \red{21.48} & \red{33.21} & \red{15.42} & \red{53.39} & \red{4.10} & \red{118.61} \\ \hline
\multicolumn{1}{|c|}{\multirow{2}{*}{TAT}}                                                    & 1,330  & 388.6    & 2,517   & 5,123   & 3,003   & 6,444   & 3.85      & 7.73           \\
\multicolumn{1}{|c|}{}                                                                        & \red{1,059} & \red{90.89} & \red{2,497} & \red{5,883}&\red{3,861} & \red{6,934} & \red{5.55} &\red{42.48} \\ \hline
\multicolumn{1}{|c|}{\multirow{2}{*}{AAT}}                                                    & 63.14    & 27.97    & 86.69     & 188.26    & 166.93    & 138.70    & 2.98     & 5.96          \\
\multicolumn{1}{|c|}{}                                                                        & \red{50.60} & \red{12.98} & \red{85.27} & \red{206.89} & \red{184.17} & \red{156.53} & \red{4.08} & \red{14.18} \\ \hline
\multicolumn{1}{|c|}{\multirow{2}{*}{\begin{tabular}[c]{@{}c@{}}TFC\\ {[}K{]}\end{tabular}}}                                                    & 7.2 & 1.7 & 15.61 & 33.5 & 17.1 & 60.10 & 4.65     & \textbf{10.05}           \\
\multicolumn{1}{|c|}{}                                                                        & \red{5.7} & \red{0.3}   & \red{15.01} & \red{38.5} & \red{20.6} & \red{88.52} & \red{6.75}     & \red{68.66} \\ \hline
\multicolumn{1}{|c|}{\multirow{2}{*}{SFC}}                                                    & 515.6    & 177.3    & 907.7     & 1,640   & 1,181   & 2,081   & 3.18     & 6.66           \\
\multicolumn{1}{|c|}{}                                                                        & \red{361.05} & \red{30.18}&\red{796.6}&\red{1,673}&\red{1,215}&\red{2,098}&\red{4.63}&\red{40.27}          \\ \hline
\multicolumn{1}{|c|}{\multirow{2}{*}{RUB}}                                                    & 0.05     & 0.00     & 0.19      & 0.07      & 0.00      & 0.22      & 1.4     & N/A           \\
\multicolumn{1}{|c|}{}                                                                        & \red{0.06} & \red{0.00}     & \red{0.22}   & \red{0.08} & \red{0.00} & \red{0.23} & \red{1.33}& \red{N/A} \\ \hline
\multicolumn{1}{|c|}{\multirow{2}{*}{RUF}}                                                    & 0.07     & 0.00     & 0.18      & 0.12      & 0.02      & 0.22      & 1.71     & N/A          \\
\multicolumn{1}{|c|}{}                                                                        & \red{0.09} & \red{0.00} & \red{0.22} & \red{0.13} & \red{0.02} & \red{0.24} & \red{1.44} & \red{N/A}  \\ \hline
\multicolumn{1}{|c|}{\multirow{2}{*}{AIT}}                                                    & 3.36     & 2.24     & 3.59      & 3.40      & 2.45      & 3.51      & 1.01     & 1.09         \\
\multicolumn{1}{|c|}{}                                                                        & \red{2.95}&\red{1.74}&\red{3.60}&\red{3.18}&\red{2.27}&\red{3.39}&\red{1.07}&\red{1.3}         \\ \hline
\multicolumn{1}{|c|}{\multirow{2}{*}{SIT}}                                                    & 5.22     & 3.44     & 5.50      & 5.14      & 3.18      & 5.28      & 0.98     & 0.92          \\
\multicolumn{1}{|c|}{}                                                                        & \red{4.09}&\red{1.98}&\red{5.06}&\red{4.72}&\red{2.79}&\red{4.96}&\red{1.15}&\red{1.41}    \\ \hline
\end{tabular}
        \vspace{-0.5cm}
\end{table}

\subsubsection{Traffic}

We extract statistical measures 
for traffic metrics (Sec. \ref{sec:traffic_analysis}) per device per day.
The \textit{CFS} algorithm was run on 19 features, reducing them to 11.
A summary of these metrics is provided in Table \ref{tab:traff_numbers}.
The correlations are depicted in Fig. \ref{fig:correlations}b.
The analysis shows us that average number of packets and bytes are positively correlated, but negatively correlated with variance of bytes and uncorrelated with IAT. 
Average IAT (\textit{AIT}) seems to be mostly independent from other traffic features, but as \textit{AIT} increases, its standard deviation (\textit{SIT}) also greatly increases which could be due to device mobility; bearing further investigation on traffic-mobility interactions. 
Interestingly, active time is \textit{weakly correlated} with number of flows and packets, which shows that users who remain online longer are \textit{not} necessarily consuming traffic at a high rate. Examining weekdays and weekends, correlation trends among traffic features remain similar for either device type.

\subsubsection{Cross-dimension}

\begin{figure}[!ht]
  \centering
  \setlength{\belowcaptionskip}{-13pt}
  \includegraphics[width=0.47\textwidth]{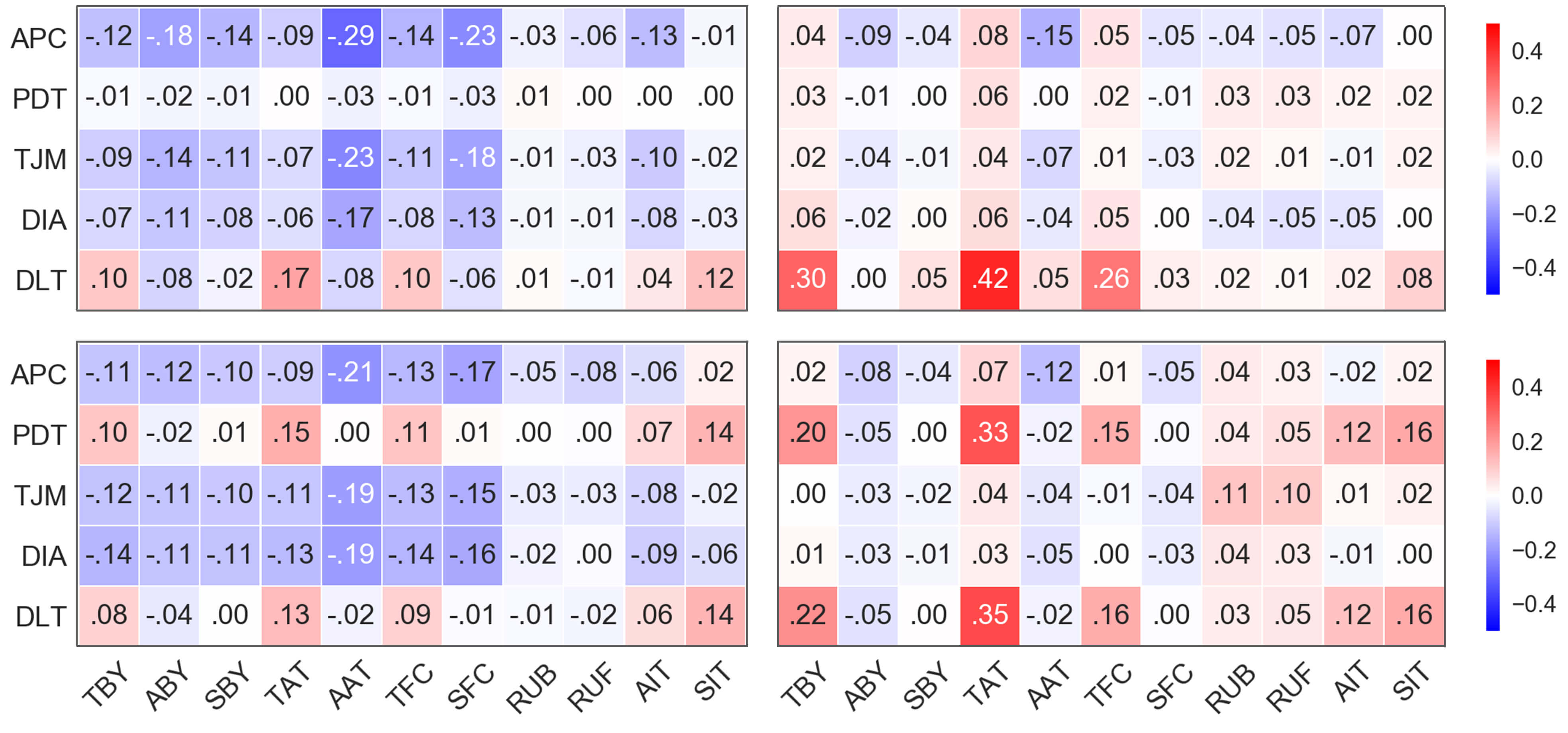}
  \caption{Correlation plots of mobility vs. traffic on weekdays (top) vs. weekends (bottom) for flutes (left) and cellos (right). }
  \label{fig:cross_corr_4in1}

\end{figure}
Studying correlations across mobility and traffic dimensions, based on subsets of features selected by \textit{CFS}, is a solid step towards an integrated mobility-traffic model. Results are presented in Fig. \ref{fig:cross_corr_4in1}.
We find that as the numbers of unique APs/buildings visited (\textit{APC}, \textit{BLD}) \textit{increase}, the average active time (\textit{AAT}), and total and std. of flow counts (\textit{TFC} and \textit{SFC}) \textit{decrease} markedly (significant negative correlation).
Surprisingly, there is no noticeable change in total traffic consumed with change in \textit{APC}, suggesting bundling of more packets in flute flows. (Similar correlation between mobility diameter and the above traffic features)
Average IAT (\textit{AIT}) of flutes also rises slightly as mobility metrics \textit{decrease}; for cellos this correlation is almost \textit{nonexistent}. 
This reinforces our \textit{``stop-to-use''} categorization; cellos are movable but are not active in transit.
To sum, \textit{flutes score high on mobility metrics}, have an overall lower flow count and network traffic but produce bigger flows on average.
For cellos, on weekends the more time spent at preferred buildings the higher the total active time (\textit{TAT}) and flow counts; this effect exists to a lesser degree for flutes. On weekdays, such correlation does not exist.

\subsection{Steps towards modeling} \label{modeling}
Here we present our steps towards an integrated mobility-traffic model, with various applications in simulation and protocol design. 
We utilize daily mobility and traffic features of users during a week.
First, we examine how different mobility and traffic features are for flutes and cellos using machine learning. 
Second, we investigate whether natural convex clusters of users appear in the dataset.
These steps verify that the differences of mobility and traffic characteristics across device types are \textit{significant}.
We also find that \textit{combining} \textit{mobility \textbf{and} traffic} makes this distinction even more \textit{clear}.
Finally, mixture models are used to model and synthesize simulated data points of each device type, finding that
the accuracy the mixture model \textit{increases} when trained on \textit{combined} features.

\begin{enumerate}[wide]
\item Supervised classification: Having shown significant differences throughout this study, we used support vector machines (SVM) 
on different subsets of features to examine the feasibility of device type inference as well as the relationship between mobility and traffic characteristics.
These sets include mobility and traffic features \textit{separately}, then \textit{combined}, and then combined with \textit{weekend/weekday labels}. 
Using \textit{solely mobility features} achieves $\approx$65\% accuracy, while \textit{traffic features alone}, obtains $\approx$79\% accuracy.
Using all mobility and traffic variables \textit{combined}, 
the trained model achieves $\approx$81\% accuracy. 
Then, as the \textbf{combined} feature set is extended to include \textit{weekdays and weekends} independently, accuracy increases to $\approx$86\%. 
This suggests that users' behavior (both flutes and cellos) is \textit{more distinguishable} when looking at \textbf{combined} mobility and traffic features; 
especially when \textit{temporal} features such as weekdays are considered separately from weekends.
We note that such behavior gaps are \textit{not} the same for both device types and a model should to take that into account.

\item Unsupervised clustering: To investigate natural convex clusters, we used K-means algorithm.
Using \textit{mobility features only}, the best mean silhouette coefficient is achieved on k=2 and 4. 
However, cluster sizes are highly skewed and at k=2, $\approx60\%$ of devices are correctly clustered.
\textit{Traffic features alone}, at k=2,  results in $\approx81.2\%$ accuracy.
\textbf{Combining} mobility and traffic, \textit{increases} the accuracy to $\approx81.5\%$. 
While some flutes and cellos are similar in terms of mobility and traffic, the clusters of the combined features clearly illustrate \textbf{two distinct modes} (especially in \textit{traffic}) and the \textit{high homogeneity} of the clusters hints at \textit{disjoint sets of behaviors} in mobility and traffic dimensions, governed by the device type.

\item Mixture model: To take a step towards synthesis of traces based on our datasets, we trained Gaussian mixture models (GMM) on \textit{combined mobility and traffic features}. From the combined model ($CM$), we acquired simulated samples. 
We used Kolmogorov-Smirnov (KS) statistic to compare the simulated samples with the real data and found that $CM$ is able to capture the behaviors of each device type.
(Average KS statistic of features is \textbf{$\approx0.15$ for flutes} and \textbf{$\approx0.14$ for cellos}. More details in appendix \ref{sec:appendix_first_model_steps}.)
Importantly,  the combined model produces samples whose \textit{traffic} features match the original data \textbf{better}, compared with training a GMM \textit{on traffic features alone}  (based on KS statistic), hinting at a key relationship between mobility and traffic. However, comparing mobility features of $CM$ with a GMM trained on mobility features alone shows no improvement.

Overall, this suggests that there is significant potential for an \textbf{\textit{integrated mobility-traffic model}} that captures the differences and \textit{\textbf{relationships}} of features, 
across \textit{\textbf{device types}, \textbf{time} and \textbf{space}}.
We leave detailed comparison of combined modeling with separate modeling of mobility and traffic for future work.
\end{enumerate}
\section{Conclusion}
\label{sec:conclusion}
In this study, we mine large-scale WLAN and NetFlow logs from a campus to answer:
\textit{
(I) How different are mobility and traffic characteristics across device types, time and space?
(II) What are the relationships between these characteristics?
(III) Should new models be devised to capture these differences? And, if so, how?}
We build \textit{FLAMeS}, a framework for systematic processing and analysis of the datasets.
Using MAC address survey, OUI matching and web domain analysis, we categorized devices: \textbf{flutes} (\textit{``on-the-go''}) and \textbf{cellos} (\textit{``stop-to-use''}).
We then study a multitude of mobility and traffic metrics, comparing flutes and cellos across time and space.
On average, flutes visit twice as many APs as cellos, while cellos generate $\approx$2x more flows.
However, flute flows are 2.5x larger in size, with $\approx$2x the number of packets. 
The best fit distribution for location preference is \textbf{Zipfian}, for flow/packet sizes is \textbf{Lognormal}, and for flow IAT at APs is \textbf{beta}.
Furthermore, flute traffic drops sharply on weekends whereas many cellos remain active.
Across mobility and traffic dimensions, we spot a negative correlation for flutes between mobility and flow duration but negligible correlation with traffic size; for cellos, this effect is less pronounced. 
We find a negative correlation with APs visited and active time, particularly for flutes.
However, no correlation exists between APs visited and traffic for cellos. 
We \textit{quantified} correlations \textit{across both mobility and traffic}.
Finally, we applied machine learning and trained a mixture model to synthesize data points and verified that the \textbf{combined} mobility-traffic features capture the \textit{differences} in metrics \textit{\textbf{better}} than \textit{either mobility or traffic separately}.
Many of our findings are not captured by today's models, and they provide insightful guidelines for the design of evaluation frameworks and simulations models.
Hence, this study answered the questions posed, introduced a strong case for newer models, and provided our first
step towards a future integrated mobility-traffic model.
\newpage
{\small
\textbf{Acknowledgments}: We thank Alin Dobra for help in the computing cluster, and the anonymous reviewers for useful feedback. 
Mostafa Ammar suggested the term \textit{'cello mobility'}.
Partial funding was provided by NSF 1320694 at Univ. of Florida, August-Wilhelm Sheer fellowship at Technical University-Munich, and Aalto Univ.
We also thank Tim Bohne of Osnabrück University for pointing out mistakes in the text.
}


\nocite{*} 

\bibliographystyle{IEEEtran}
\bibliography{main_arXiv}


\numberwithin{equation}{section}

\section*{Appendices}
\setcounter{section}{0}

Here we further describe various aspects of our submitted work to Infocom 2018,
which were not included in the original document for brevity. 

\section{Merging Datasets}
\label{sec:appendix_merge}
\begin{figure}[ht]
	\centering
	\setlength{\belowcaptionskip}{-5pt}
	\includegraphics[width=0.45\textwidth]{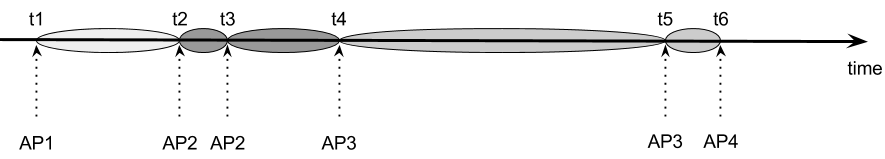}
	\caption{Wireless association for a device at different times.}
	\label{fig:dhcptime}
\end{figure}

In order to study network traffic across devices and APs, it is necessary to match the NetFlow
records with wireless associations (from WLAN dataset). This task requires the MAC-IP mapping.
The IP addresses are dynamically assigned using DHCP but DHCP session logs were not directly available and had to be derived. 
We define the duration of a DHCP lease as the time between two consecutive associations of a device with any AP; i.e. when a device connects to $AP_1$, a session starts and once the user device connects to $AP_2$, the first session ends and a new one starts. 
Fig. \ref{fig:dhcptime} illustrates the associations of a sample device with different APs at different times. 
The first session would have the IP given by $AP_1$ and a lease time $t_2 - t_1$, and so on. (total of 5 sessions in this example)
The last association is discarded as we do not know the duration of that IP assignment.
Combining these derived-DHCP records with the \textit{Location Information} and \textit{Device Type Classification} we create the \textbf{DHCP} table.

The derived DHCP and NetFlow datasets were then merged to form what we refer to as the \textbf{CORE} dataset for our study. The unique identifiers between the two are the clients' IPs in addition to start and end time of flows, hence the need for a DHCP-like set. For a DHCP lease session \textit{LS}, all flows whose IP address is the same as the lease \textit{and} whose entire lifetime falls within the lease duration, are associated with \textit{LS}.

Given these traces, cellular usage cannot be analyzed. However, this does not significantly impact analysis for two reasons:
1) The traces already capture a very large user-base, with tens of thousands of active devices. This raises confidence in our analysis of a real world WLAN. 
2) The WiFi campus coverage is ubiquitous, with 1760 APs installed in the vast majority of populated areas. Also, most laptops on campus lack cellular connectivity, and many smartphones use WiFi for their data to avoid cellular data costs.

\section{Computing System} \label{sec:appendix_computing_sys}

The size of the datasets is $\approx$30TB in raw text format, mostly
consisting of NetFlow data and $\approx$0.5TB for AP logs. There
were several challenges in managing and mining the largescale
datasets that required a thorough preparation, to run on
a fast machine with plenty of resources/memory.
We explored several techniques and pipelines for extraction,
transformation, loading (ETL) and querying of big data and
chose tools from Apache Hadoop ecosystem. We use Hive as
our data warehouse (tables stored in Parquet format). Apache
Spark is the compute engine for data processing and analysis
tasks. Computation runs on two nodes, each with 64 cores
and $\approx$0.5TB of memory. Further discussion of the system
and comparison to others is out of scope of this document.

\section{Mobility Analysis} \label{sec:appendix_mobility}

For completeness, we include further analysis of the mobility aspects of our dataset, discussed in Section \ref{sec:mobility}
of the conference paper.

\begin{figure}[h]
	\centering
	\setlength{\belowcaptionskip}{-5pt}
	\begin{subfigure}[b]{0.48\linewidth}
		\includegraphics[width=\textwidth]{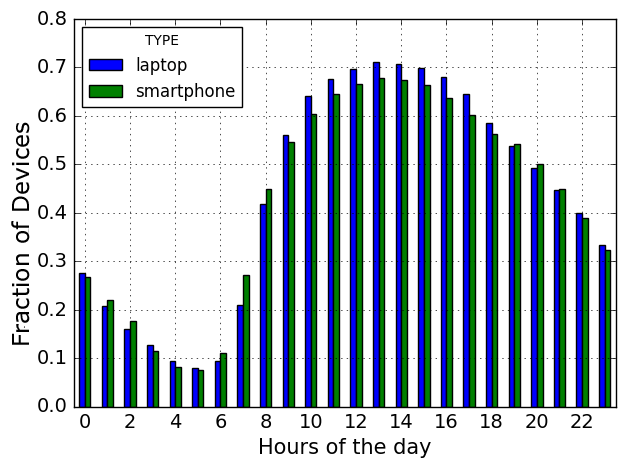}
		\caption{}
		\label{fig:sfl}
	\end{subfigure}
	\begin{subfigure}[b]{0.50\linewidth}
		\includegraphics[width=\textwidth]{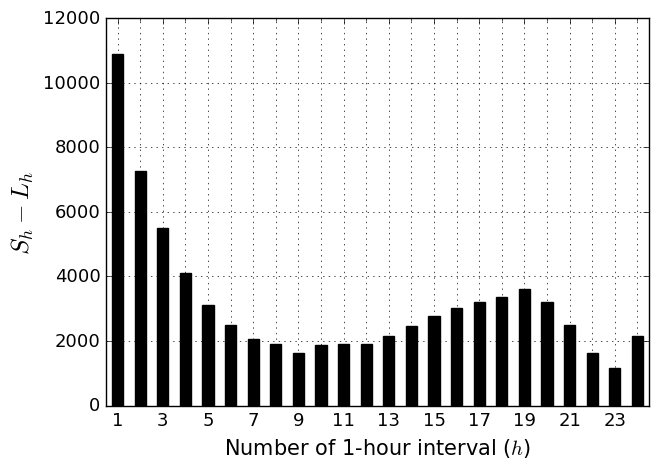}
		\caption{}
		\label{fig:sfl_diff}
	\end{subfigure}
	\caption{Hourly associations.}
	\label{fig:sfl_both}
\end{figure}

\subsection{Hourly associations}
Measuring device associations every hour, Fig. \ref{fig:sfl} shows the percentage of devices with at least one event as a function of hours of the day.
The majority of devices appear online between 9am and 8pm, with the hours between 2am and 6am having less than 20\% of devices associating. 
We find no major differences between flutes' and cellos' distributions, as many users potentially own both. 
As users arrive on campus and their phones announce their first location, they switch on their laptops. 
This issue bears further research through a future census study.

To measure the stay of devices throughout a day, we look at 1-hour intervals, 
and measure the number of hours a device accessed an AP \footnote{ P. Widhalm, et al., “Discovering urban activity patterns in cell phone data”.}.
Fig. \ref{fig:sfl_diff} depicts $S_h - L_h$, where $S_h$ and $L_h$ are total number of flutes and cellos respectively,
with at least one record per hour, as a function of the number of hours online $h$.
Flutes are predominant for short visits and very long stays, but the difference drops significantly at 9 hours, then increases. 
The rise after 9 hours is likely due to students living on campus, with always-on connected phones.

\subsection{Visitation preferences}

\begin{figure}[h]
	\centering
	\setlength{\belowcaptionskip}{-6pt}
	\begin{subfigure}[b]{0.4\linewidth}
		\includegraphics[width=\textwidth]{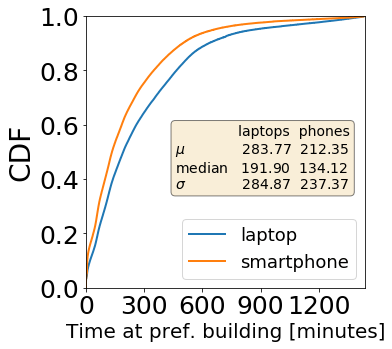}
		\caption{}
		\label{fig:t_pref_bld}
	\end{subfigure}
	\begin{subfigure}[b]{0.58\linewidth}
		\includegraphics[width=\textwidth]{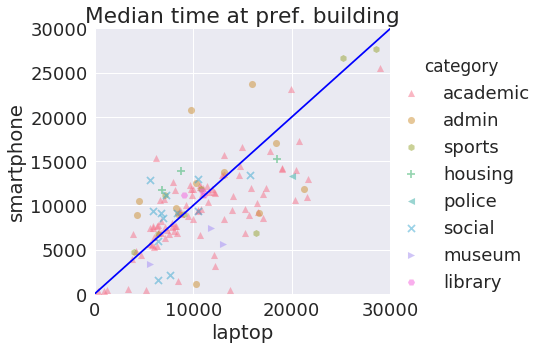}
		\caption{}
		\label{fig:t_pref_scatt}
	\end{subfigure}
	\caption{Time spent at preferred building.}
	\label{fig:timeprefbld}
\end{figure}

Fig. \ref{fig:t_pref_scatt} shows a scatter of the median time spent at a user's preferred building.
Each dot represents this value for a given location.
This plot shows that \textit{academic}, \textit{police} and \textit{museum} buildings tend to have laptops staying longer,
which makes sense intuitively, with students using laptops during lectures and staff working at the other two categories.
On the contrary, for \textit{social} and \textit{housing} buildings, there is a higher probability of having flutes staying longer,
hinting at a tendency to use mobile devices more in such places.
Finally, \textit{administrative}, \textit{sports} and \textit{library} buildings tend to have both types of devices staying for similar amounts of time. 
Analysis of inherited differences in browsing of online services given by this heterogeneity among buildings is left for future work.

Fig. \ref{fig:t_pref_bld} depicts the time devices spend at their \textit{preferred building} in a day.

\subsection{Return probability}

\begin{figure}[ht]
	\centering
	\setlength{\belowcaptionskip}{-10pt}
	\includegraphics[width=0.68\linewidth]{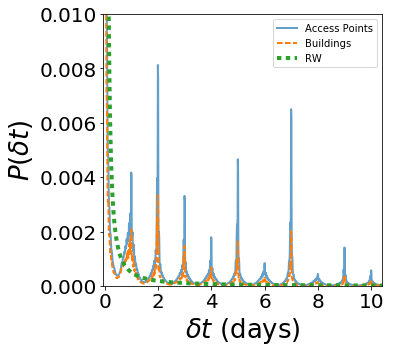}
	\caption{Probability to return to a previously visited location.}
	\label{fig:returnp}
\end{figure}

We compare empirical values for devices to return to previously visited APs or buildings in Fig. \ref{fig:returnp}.
We observe returning spikes at every 24 hours, with the highest peaks at 48 and 168 hours (2 and 7 days).
This can be explained by the schedule of classes at the university.

\section{Traffic Analysis} \label{sec:appendix_traffic}

In this Section, we further discuss references from Section \ref{sec:traffic_analysis} of
the conference paper.

\subsection{Flow sizes}
 This metric is the sum of bytes for all packets
within a single flow. First, outlier data points are removed
using a robust measure of scale, based on inter-quartile range
(IQR). Looking
at individual flows of each device type shows that size of flows
that originated from smartphones are significantly different
that laptop flows (p-value< .05).
\footnote{Flow metrics do not fit Gaussian distribution (based on Shapiro-Wilk test for
	normality, goodness-of-fit test and Q-Q plot results, not included for brevity).
	Thus, we use Mann-Whitney statistical test\footnote{Henry B Mann et al, "On a test of whether one
		of two random variables is stochastically larger than the other. The
		annals of mathematical statistics".} to compare two unpaired
	groups (laptops vs smartphones), and Wilcoxon signed-rank test to compare
	two paired groups (each device type on weekdays vs weekends) \footnote{ Frank Wilcoxon, "Individual comparisons by ranking methods".}.}

\begin{figure}[ht]
	\centering
	\setlength{\belowcaptionskip}{-14pt}
	\includegraphics[width=0.85\linewidth]{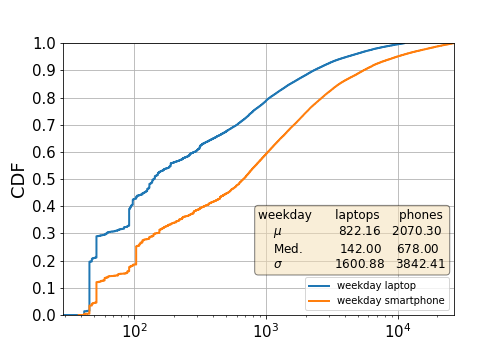}
	\caption{CDF of individual flow sizes (bytes, log-scale $x$ axis), similar pattern on weekends}
	\label{fig:flowsizes}
\end{figure}

On weekdays, the average
size for smartphone flows is 2070 bytes and 822 bytes for
laptop flows; with no significant changes on weekends (CDF
in Fig. \ref{fig:flowsizes}). The difference in medians is more pronounced,
on weekdays, for smartphones it is 678 bytes while it is 142
bytes for laptops (similar values on weekends).


\subsection{Lognormal plots}

\begin{figure}[ht]
	\centering
	\setlength{\belowcaptionskip}{-14pt}
	\includegraphics[width=0.85\linewidth]{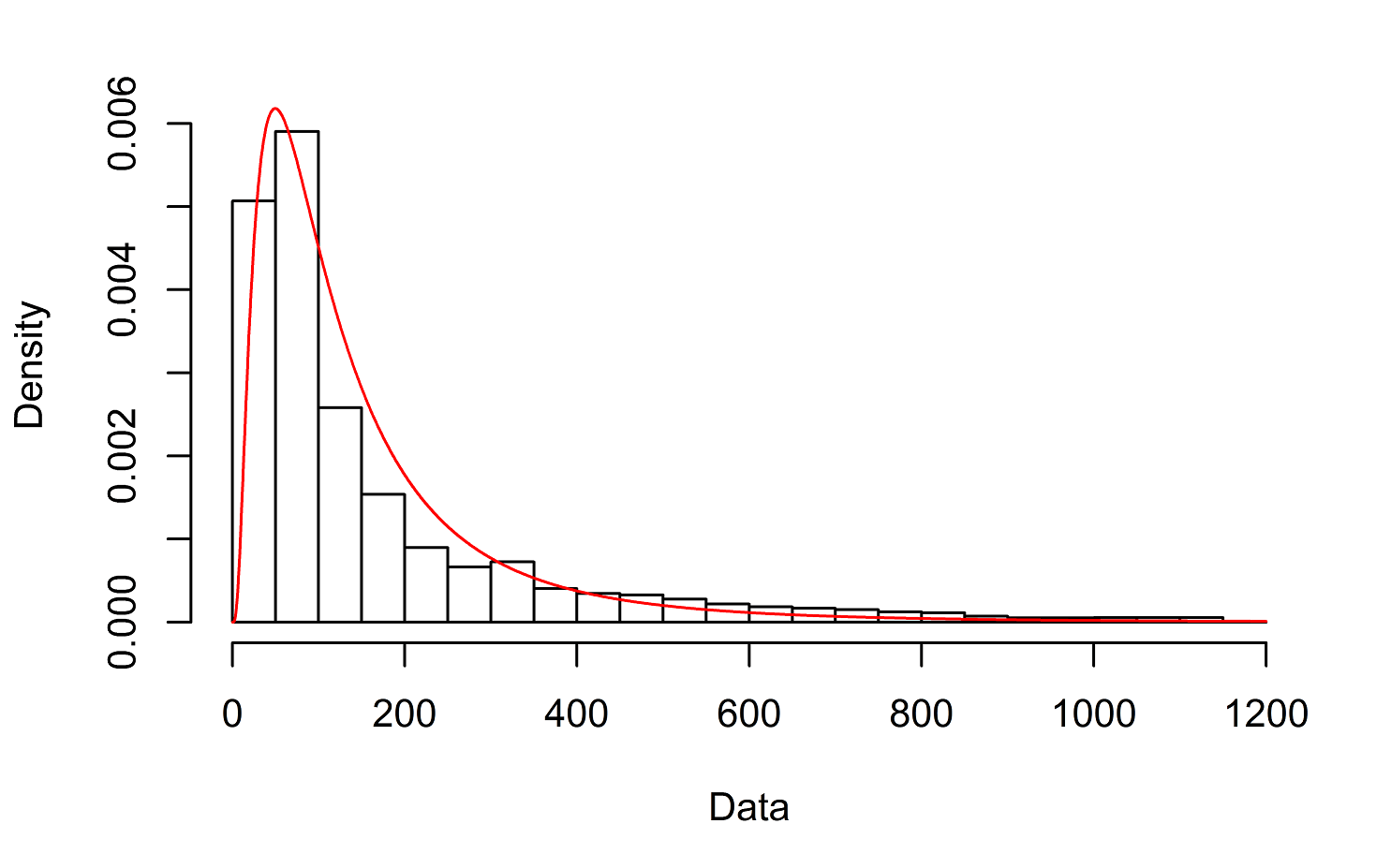}
	\caption{Lognormal distribution.}
	\label{fig:lognormal}
\end{figure}

For flow sizes in our dataset, a Lognormal distribution is
the best fit, regardless of device type (Fig. \ref{fig:lognormal}). Many models
assume flow sizes are static, or follow an exponential distribution
but real world data provides no supporting evidence.
Such simplifying assumptions fail to accurately account for
very large flows obtained from a Lognormal distribution.

\subsection{beta distribution}

\begin{figure}[ht]
	\centering
	\includegraphics[width=0.9\linewidth]{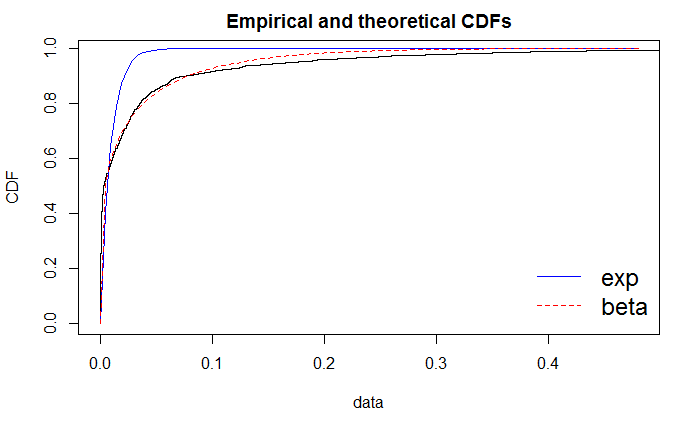}
	\caption{Exponential and Beta distributions.}
	\label{fig:expbeta}
\end{figure}

Inter-arrival times (IAT). Our results show that
the flow IAT, regardless of device type, does not follow an

exponential distribution. Flow IAT matches a beta distribution
well (Fig. \ref{fig:expbeta}) with a very high estimated kurtosis and
skewness (estimated at 58 \& 6.9 respectively). The high estimated
kurtosis illustrates that there are infrequent extreme
values, which explains the observed highly elevated standard
deviation of IAT\footnote{In the research community, packet IAT and its Fourier transform are considered
	important features in traffic analysis. They are used extensively in
	simulation and modeling of networking protocols as well as internet traffic
	classification \cite{Falaki2010}. Realistic modeling of IAT is required for accurate simulation
	and measurement of congestion control mechanisms \cite{Moore2005}. Due to
	limited availability or staleness of most packet-level datasets, although our
	NetFlow is on a higher abstraction layer (flow-level vs packet-level), analysis
	of flow IAT can still be used for measuring delay and jitter effects.}

\section{First Modeling Steps}
\label{sec:appendix_first_model_steps}
\begin{figure}[ht]
	\centering
	\setlength{\belowcaptionskip}{-14pt}
	\includegraphics[width=0.9\linewidth]{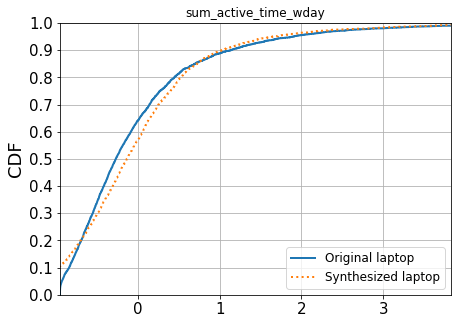}
	\caption{Synthetic vs. Original \textit{TAT} feature for flutes.}
	\label{fig:synthetic}
\end{figure}

More details of the KS test and GMM model are provided here.
We found that providing both mobility and traffic features to train a GMM results in lower average KS statistic.
Fig \ref{fig:synthetic} shows a sample CDF of \textit{TAT}. KS statistic details can be found in Table \ref{tab:ks}.

\begin{table}[]
\centering
\caption{KS-statistic summary}
\label{tab:ks}
\begin{tabular}{cccll}
\hline
KS statistic & Flutes & Cellos &  &  \\ \hline
Average      & 0.150  & 0.140  &  &  \\
Min          & 0.052  & 0.027  &  &  \\
Max          & 0.380  & 0.350  &  &  \\
Std          & 0.086  & 0.0787 &  & 
\end{tabular}
\end{table}

\section{Lessons Learned and Modeling Insights}
\label{sec:appendix_modeling}
Our above findings provide further (but surely not yet comprehensive)
insights into considerations relevant to the design and
parameterization of mobility and network traffic models.  While we
leave devising and validating a concrete candidate model for future
work, we can readily identify the following important elements:

It is crucial to differentiate flutes vs. cellos for both mobility and
traffic due to their very different nature.  More specifically, flutes
exhibit continuous presence whereas cellos are on/off with jumps
between locations.  Beyond differences in continuity, the traffic patterns
(flow sizes, arrival times, etc.) should be specified by device class.
Moreover, the traffic generation, spatial locations, and temporal behavior
can be linked per device type and per user ``community'' (e.g. students
of different disciplines at various buildings).

Our analyses allow us to quantify the correlated elements of traffic
and motion for the (type of) campus we investigated.  AP associations
allow us to calibrate user mobility and determine their community
structures (and ``hotspots''), while daily and weekly activity
patterns help outline user schedules---which could be parameterized
based upon outline information such as class schedules, public
holidays, and other predictable events.  Traffic flows per device can
be created from our corresponding distributions, while AP traffic
observations allow parameterizing the above ``hotspots'' with respect
to user activity.

Together, these could form a basis for generating parameters for
established mobility models\footnote{Hsu, W-J., et al. "Modeling time-variant user mobility in wireless mobile networks." INFOCOM 2007}\textsuperscript{,}\footnote{Boldrini, Chiara, and Andrea Passarella. "HCMM: Modelling spatial and temporal properties of human mobility driven by users’ social relationships." Computer Communications 2010},
possibly extend them as needed, and augment them
with adequate correlated traffic models.  The next step is examining
the minimum number of parameters required (parsimonious model) to
approximate user behavior within given error bounds with respect to
the observed ground truth.
%
This should be complemented by studying datasets from other network
settings to further validate the observed behaviors beyond our
campus environment, taking into account the specifics of the campus and
identifying further influencing factors.

Besides different campus environments, a few further elements deserve
further study: 1) Our study so far gives only limited insight into the
actual services users are accessing, so analyzing web domains
users are accessing, for interest and application analysis as well as
investigating the relationship of user interests with spatio-temporal
characteristics of mobility and traffic would be important next steps.
2) Flutes and cellos are likely two points in a dimension that is
continuously changing as a) differentiation in device classes appear
(e.g., smartphones vs tablets---do we have ``guitars'' as well?); b)
blur again (e.g., due to new form factors for smartphones and
tablets); and c) new devices such as fitness trackers, smart watches
and ``glasses'' further enrich the portfolio.

\end{document}